\pdfoutput=1
\documentclass{article}
\usepackage{arxiv}
\usepackage[numbers,sort&compress]{natbib}
\usepackage[utf8]{inputenc}
\usepackage[T1]{fontenc}
\usepackage{lmodern}
\usepackage[expansion=false]{microtype}
\usepackage{amsmath,amssymb,amsthm}
\usepackage{array}
\usepackage{booktabs}
\usepackage{graphicx}
\usepackage{caption}
\usepackage{enumitem}
\usepackage[section]{placeins}
\usepackage[hyphens]{url}
\usepackage[hidelinks]{hyperref}
\hypersetup{pdftitle={Reliability-Aware Z-Grammars and Z-Automata over a Credible-Grade Semiring},pdfauthor={Sherzod Turaev, Nor Haniza Sarmin, Wan Heng Fong},pdfkeywords={Z-numbers; weighted automata; fuzzy grammars; selective semiring; reliability-aware parsing; formal power series}}

\theoremstyle{plain}
\newtheorem{theorem}{Theorem}[section]
\newtheorem{proposition}[theorem]{Proposition}
\newtheorem{corollary}[theorem]{Corollary}
\theoremstyle{definition}
\newtheorem{definition}[theorem]{Definition}
\newtheorem{example}[theorem]{Example}
\theoremstyle{remark}
\newtheorem{remark}[theorem]{Remark}

\newcommand{\Sem}{\mathbb{K}}

\newcommand{\Real}{\mathbb{R}}
\newcommand{\acut}[1]{{}^{#1}\!}
\newcommand{\sem}[1]{[\![#1]\!]}
\newcommand{\Zset}{\mathsf{Z}}
\newcommand{\Zz}{\mathcal{Z}_{0}}
\newcommand{\Zzz}{\mathcal{Z}_{00}}
\newcommand{\Zlang}{\mathcal{L}}
\newcommand{\beh}[1]{\lVert#1\rVert}
\newcommand{\grade}{\pi_{A}}
\newcommand{\rel}{\pi_{B}}
\newcommand{\cred}{\delta}
\newcommand{\otimesz}{\otimes}
\newcommand{\oplusz}{\oplus}
\newcommand{\series}[1]{#1\langle\!\langle\Sigma^{*}\rangle\!\rangle}
\newcommand{\rec}{\mathrm{Rec}}
\newcommand{\rat}{\mathrm{Rat}}
\newcommand{\Vit}{\mathbb{V}}

\title{Reliability-Aware Z-Grammars and Z-Automata over a Credible-Grade Semiring}

\author{
  Sherzod Turaev\thanks{Corresponding author. ORCID: 0000-0001-6661-8469.} \\
  Department of Computer Science and Software Engineering \\
  College of Computing and Artificial Intelligence \\
  United Arab Emirates University \\
  Al Ain, United Arab Emirates \\
  \texttt{sherzod@uaeu.ac.ae} \\
  \And
  Nor Haniza Sarmin\thanks{ORCID: 0000-0003-4291-5746.} \\
  Department of Mathematical Sciences, Faculty of Science \\
  Universiti Teknologi Malaysia \\
  Johor Bahru, Malaysia \\
  \texttt{nhs@utm.my} \\
  \And
  Wan Heng Fong\thanks{ORCID: 0000-0001-8202-4388.} \\
  Department of Mathematical Sciences, Faculty of Science \\
  Universiti Teknologi Malaysia \\
  Johor Bahru, Malaysia \\
  \texttt{fwh@utm.my} \\
}
\date{}

\begin{document}

\maketitle

\begin{abstract}
Fuzzy grammars and weighted automata usually attach to each word a single grade, which records how strongly the word is generated but not how far that record can be trusted. We study weighted grammars and finite automata whose weights are scalar grade-reliability pairs inspired by Z-numbers. Sequential composition multiplies both components, while aggregation selects a derivation by its credible grade, the product of grade and reliability, with ties resolved by grade. On the canonical carrier, in which the weights of zero grade or zero reliability are identified with zero, the resulting structure is a commutative selective semiring, isomorphic to a subsemiring of a lexicographically ordered product, and its credible-grade projection is a homomorphism onto the Viterbi semiring. We compare several alternative combination rules and identify the carriers on which the relevant algebraic laws hold. Standard results on weighted automata then yield a rational characterization and closure properties. We prove that positive credible-grade cuts and selected-run reliability filters of finite automata are regular and effectively constructible for rational weights and thresholds, and we exhibit distinctions that the scalar projection cannot see: a reliability filter that no scalar reweighting of the same one-state automaton implements, nonmonotonicity under the addition of runs, and an unambiguous automaton with no deterministic equivalent although its scalar projection is deterministic. A count-based example illustrates an attestation penalty for grammar rules. The construction concerns scalar pairs, and its extension to arbitrary fuzzy-valued Z-numbers remains open.
\end{abstract}


\section{Introduction}\label{sec:intro}

The theory of formal languages has, from early in its history, been generalized so that a grammar or an automaton carries not only the bare fact that a word belongs to a language but a quantity attached to that word. The fuzzy grammars of Lee and Zadeh~\cite{lee1969} graded each production by a membership degree and read the membership of a word as the best such degree along a derivation. The weighted automata that Sch\"utzenberger~\cite{schutz1961} introduced in the same period assigned to each transition an element of a semiring and computed the behavior of the automaton as a formal power series. The modern synthesis collected these and many other developments into a single theory of weighted automata over an arbitrary semiring, within which fuzzy, probabilistic, and distance-based devices are recovered by a choice of the weight structure~\cite{dkv2009}. In the fuzzy and probabilistic instances that motivate the present work, the weight is a single number, whether a membership grade, a probability, a cost, or a value in a lattice-ordered monoid~\cite{li2005}, and it records how strongly a word is generated. Multicomponent weights are not new in themselves, since product, lexicographic, and derivation-carrying semirings attach several quantities to a path~\cite{goodman1999,roark2011}, and intuitionistic fuzzy automata carry a degree of membership together with a degree of nonmembership~\cite{jun2005}. In none of these, however, does the second component qualify how far the first can be trusted.

A single number is not always an adequate record of what is known about a quantity. In the elicited judgments on which decision models increasingly rely, and in the estimates that data-driven systems derive from finite samples, a value arrives together with an indication of how far it is to be trusted, and the two should not be collapsed into one. Zadeh~\cite{zadeh2011} proposed the \emph{Z-number} as the formal device for exactly this pairing. A Z-number $Z=(A,B)$ attaches to a variable a fuzzy restriction $A$, the imprecise value itself, together with a second fuzzy quantity $B$ that measures the \emph{reliability} of that restriction, understood as a graded constraint on the probability that the variable indeed satisfies $A$. The construction has been developed into a full arithmetic and a decision calculus~\cite{aliev2015,aliev2016} and applied across the quantitative decision sciences. Our own work on sustainable trip planning has used it to attach to the rating of a place a reliability built from the number of reviews behind it, benchmarking that evidence-aware representation against alternative formalisms across a national catalog~\cite{turaev2026}. In all of this the Z-number is a datum or a decision variable; it has not been asked to pass through a grammar.

The question this paper takes up is whether it can. Formal language theory knows how to carry a grade through a derivation, and the second component of a Z-number is, on its face, just another number in $[0,1]$, so one might expect the reliability to ride along beside the grade, each component combined by its own operations as a derivation proceeds. The expectation is sound algebraically and unsound semantically, and the gap between the two is instructive rather than incidental. If grades and reliabilities are each composed along a derivation and each aggregated across competing derivations by their own operations, the result is a product of two semirings and hence a semiring. The pair it reports for an ambiguous word, however, may join the best grade of one derivation to the best reliability of another, and then it describes no derivation of the word at all. A reliability qualifies a particular derivation and must stay attached to it, which requires \emph{coupling} the two components: the competing derivations are compared, one of them is selected, and its grade and its reliability are reported together. Coupling by selection keeps the reported pair coherent, but it puts the semiring laws at risk, since distributivity then fails at the \emph{degenerate} weights, in which one component is zero and the other is not. A structure that is not a semiring may lose the agreement of the recognizable and the rational series, the agreement of the run and the initial-algebra semantics, and the decision procedures that make the theory of weighted automata usable, so a naively coupled Z-grammar would be a definition without a theory. Carrying a reliability through a grammar is therefore not a matter of pairing two known constructions. It requires a coupled combination of the two components under which the weights still form a semiring, together with an argument that this combination is faithful to what the reliability is supposed to mean.

The existing generalizations of fuzzy grammars do not address this, because they generalize a different part of the device. The computing-with-words automata of Ying~\cite{ying2002}, extended to Turing machines by Wang and Qiu~\cite{wang2003}, replace the \emph{input} symbols of a fuzzy automaton by words in the sense of linguistic terms, so that the machine reads imprecise observations, while the transition weight remains a scalar grade; the generalization is on the alphabet, not on the weight. Grammars built on type-2 fuzzy sets, whose theory goes back to Mizumoto and Tanaka~\cite{mizumoto1976}, replace the grade itself by a type-2 membership, a fuzzy set of grades, which is a second order of fuzziness about the value, whereas a Z-number's second component is a reliability about the probability of the first, a different kind of second order altogether. Neither development carries a reliability through a derivation, and neither meets the coupling problem just described, since in neither must a second component stay attached to the derivation it qualifies. The multicomponent semirings mentioned above do carry a second quantity, but it either describes the word on the same footing as the first or merely breaks ties (Section~\ref{ssec:cw}). To the best of our knowledge, no weighted grammar or automaton whose weights carry the reliability component of a Z-number has been proposed, and the extensions of Z-numbers surveyed by Banerjee, Pal, and Pal~\cite{banerjee2022} include no formal-language model. The construction of such devices is the contribution of this work.

We resolve the problem as follows. We define the Z-grammar and the Z-automaton as weighted devices whose weights are \emph{Z-weights}, pairs of a grade and its reliability, combined by a weight structure that fixes how the two components compose along a derivation and how competing derivations are aggregated. We study scalar grade-reliability pairs inspired by Z-numbers. The semiring developed below is a model for these pairs; it is not a semiring of arbitrary fuzzy-valued Z-numbers, and no general arithmetic-preserving defuzzification is assumed, so that the fuzzy-valued extension is left open (Remark~\ref{rem:fuzzyvalued}). Along a derivation, grades and reliabilities multiply. The multiplicative law for reliabilities admits a probabilistic reading: if the scalar reliability $b$ of a restriction is read as a probability $b=P(X\text{ is }A)$ that the restriction holds, then under an explicit independence assumption between occurrences the probability that all the restrictions used by a derivation hold is the product of their reliabilities (Section~2). Where that assumption cannot be defended, notably when one estimated rule is used more than once, we read the product as a declared path score. Across derivations, the structure selects the derivation of greatest \emph{credible grade}, the grade discounted by its reliability, and reports its grade and its reliability together. We prove that this \emph{reliability-aware credible-grade} structure, once its degenerate weights are identified with the zero of the semiring, is a commutative semiring whose credible-grade map is a homomorphism onto the Viterbi semiring. The Z-grammars and Z-automata built over it therefore inherit the theory of weighted automata over a commutative semiring, now carrying a reliability alongside the grade at every step. We then compare it with alternative structures obtained from the same parameter families, and we illustrate with a complete count-based example how the reliability can act as an attestation penalty on grammar rules.

The specific contributions of the paper are the following.
\begin{enumerate}[leftmargin=*,itemsep=0.35ex,topsep=0.4ex]
\item We isolate what the reliability component adds beyond its scalar projection, which is where the results specific to the pair lie. A reliability filter on a one-state Z-automaton cannot be implemented by any positive scalar reweighting of the same skeleton (Proposition~\ref{prop:notscalar}); the filter on the selected run is not monotone under the addition of runs, whereas every upward scalar threshold is (Proposition~\ref{prop:nonmono}); and an unambiguous Z-automaton can have no deterministic equivalent over any commutative semiring containing the Z-weights, although its scalar projection is deterministic (Proposition~\ref{prop:nondet}). We also prove that the positive credible-grade cuts and the selected-run reliability filters of every finite Z-automaton with rational weights and thresholds are regular and effectively constructible (Proposition~\ref{prop:cut}).
\item We define the \emph{Z-grammar} and the \emph{Z-automaton} over grade-reliability pairs and a \emph{coupled} combination of the two components, the credible-grade structure $\Zset$, which multiplies both components along a derivation and selects among derivations by the credible grade, ties broken by grade. On the canonical carrier $\{\mathbf{0}\}\cup(0,1]^{2}$ it is a commutative selective semiring, the map $(a,b)\mapsto(ab,a)$ embeds it into the lexicographic product of two Viterbi semirings, and the credible grade is a semiring homomorphism onto the Viterbi semiring. Neither the embedding nor the semiring laws are new as general theorems, since selective semirings and lexicographic products are classical; what is specific is the choice of the order and of the carrier.
\item We compare alternative combination rules through a three-parameter family of \emph{Z-weight schemes}, fixed by a grade law, a path-reliability law, and an aggregation scheme, and identify the carriers on which eight representative structures from these families satisfy the semiring laws. It shows that the separable structure is a semiring on the full carrier but decouples the reported reliability from the selected derivation, that select-by-grade is a semiring on the grade-zero quotient but induces no operation on the double quotient, and that coupling by selection fails distributivity on every carrier examined when grades are combined by the minimum.
\item We record the \emph{rational theory} of Z-languages that follows from the semiring laws, namely the Kleene--Sch\"utzenberger characterization, the conservative extension of the max-product fuzzy theory, the closure properties, and the regularity of supports; these are standard corollaries and are labeled as such. We prove that the twins property suffices for the determinization of unambiguous Z-automata, leave the ambiguous case open, and locate the boundary at which behavioral equivalence becomes undecidable.
\item We give a complete \emph{count-based example} in which the grade of a rule is its maximum-likelihood probability and its reliability is an evidence score that grows with its count. The example shows that this attestation penalty can change the selected analysis; it is not an evaluation of parsing accuracy.
\end{enumerate}

The classification rests on a deliberate methodological choice. Each positive semiring claim is proved in the text, and each distributivity failure among the scalar structures is witnessed by an explicit counterexample, found by exact rational computation over a fixed grid of grades and reliabilities. The grid computation proves no semiring law, since a finite grid cannot; it locates the failures and confirms the proved identities at the grid points, so that the boundary between the semirings and the strong bimonoids is drawn by proofs and checked witnesses rather than by expectation. Two structures that a quick reading would treat alike are separated in this way. The separable structure is a semiring on the full carrier, whereas the select-by-grade structure, which couples the two components, fails distributivity at the weights of grade zero and nonzero reliability and becomes a semiring only on a quotient. The reliability-aware structure is singled out because it is a semiring that keeps the reported pair attached to a single derivation and lets the reliability enter the choice among derivations, not because it was preferred in advance.

The paper is organized as follows. Section~2 recalls fuzzy numbers, Z-numbers, and the background on grammars and weighted automata, and it separates the three readings of reliability that the paper keeps apart. Section~3 defines the Z-weight structure, the Z-grammar, and the Z-automaton and proves the semiring laws. Section~4 compares alternative combination rules and the carriers on which they satisfy these laws. Section~5 develops the rational theory and the results specific to the reliability component, Section~6 gives the count-based example, and Section~7 concludes with the limitations and the open problems.

\FloatBarrier
\section{Preliminaries}

The construction developed in the sequel draws on three bodies of theory that are rarely brought into contact, namely the theory of \emph{fuzzy sets} and the \emph{Z-numbers} built upon them, the classical theory of \emph{formal grammars} and \emph{automata}, and the theory of \emph{weighted automata} over algebraic weight structures. This section fixes the notation and the facts from each that are needed later, keeps the exposition self-contained at the level of a reader familiar with one of the three, and illustrates every notion with a worked example, so that the definitions of a Z-grammar and a Z-automaton introduced afterwards rest on ground that has already been made concrete. We refer the reader to Zadeh~\cite{zadeh1965,zadeh2011}, Aliev et al.~\cite{aliev2015,aliev2016}, Hopcroft and Ullman~\cite{hopcroft1979}, Mordeson and Malik~\cite{mordeson2002}, and Droste, Kuich, and Vogler~\cite{dkv2009} for the comprehensive treatments that these preliminaries only summarize.

\subsection{Fuzzy sets, fuzzy numbers, and their operations}

A \emph{fuzzy set} $A$ on a universe $X$ is identified with its \emph{membership function} $\mu_A\colon X\to[0,1]$, where $\mu_A(x)$ is read as the degree to which $x$ belongs to $A$, so that the ordinary (crisp) subsets of $X$ are recovered as the fuzzy sets whose membership takes only the values $0$ and $1$. The \emph{support} of $A$ is $\operatorname{supp}(A)=\{x:\mu_A(x)>0\}$, its \emph{core} is $\operatorname{core}(A)=\{x:\mu_A(x)=1\}$, its \emph{height} is $\sup_x\mu_A(x)$, and $A$ is said to be \emph{normal} when its height equals one. For a level $\alpha\in(0,1]$, the \emph{$\alpha$-cut} of $A$ is the crisp set $\acut{\alpha}A=\{x:\mu_A(x)\ge\alpha\}$, and the family of $\alpha$-cuts determines $A$ uniquely through $\mu_A(x)=\sup\{\alpha:x\in\acut{\alpha}A\}$, which is the device by which operations on fuzzy quantities are reduced to operations on ordinary intervals.

A \emph{fuzzy number} is a fuzzy set $A$ on the real line that is normal, whose $\alpha$-cuts for $\alpha\in(0,1]$ are closed bounded intervals $\acut{\alpha}A=[\underline{a}(\alpha),\overline{a}(\alpha)]$, and whose support has compact closure; equivalently, $A$ is normal, convex, and upper semicontinuous with bounded support. Boundedness of the support does not follow from the compactness of the $\alpha$-cuts, since the membership function $e^{-x^{2}}$ has the compact $\alpha$-cuts $[-\sqrt{-\ln\alpha},\sqrt{-\ln\alpha}]$ for every $\alpha\in(0,1]$ and an unbounded support. Two parametric families suffice for almost all of the examples below. A \emph{triangular} fuzzy number, written $A=(a_1,a_2,a_3)$ with $a_1\le a_2\le a_3$, has membership rising linearly from $0$ at $a_1$ to $1$ at the \emph{peak} $a_2$ and falling linearly back to $0$ at $a_3$; a \emph{trapezoidal} fuzzy number $A=(a_1,a_2,a_3,a_4)$ is flat at height one on the whole core $[a_2,a_3]$. The arithmetic of fuzzy numbers is governed by \emph{Zadeh's extension principle}, under which a function $f\colon\Real^n\to\Real$ is lifted to fuzzy arguments by
\begin{equation}\label{eq:ext}
\mu_{f(A_1,\dots,A_n)}(y)=\sup_{\,y=f(x_1,\dots,x_n)}\ \min_{1\le i\le n}\mu_{A_i}(x_i),
\end{equation}
and which, for addition and for multiplication by a positive scalar, acts cut by cut as ordinary interval arithmetic, $\acut{\alpha}(A\oplus B)=[\underline{a}(\alpha)+\underline{b}(\alpha),\ \overline{a}(\alpha)+\overline{b}(\alpha)]$.

\begin{example}[A triangular fuzzy number and a sum]\label{ex:tri}
Let $A=(2,3,4)$ represent the imprecise quantity \emph{about} $3$. Its membership is $\mu_A(x)=x-2$ on $[2,3]$, $\mu_A(x)=4-x$ on $[3,4]$, and zero elsewhere, and its $\alpha$-cut is the interval $\acut{\alpha}A=[2+\alpha,\,4-\alpha]$. Adding a second triangular number $B=(1,2,4)$, the extension principle gives, level by level, $\acut{\alpha}(A\oplus B)=[(2+\alpha)+(1+\alpha),\,(4-\alpha)+(4-2\alpha)]=[3+2\alpha,\,8-3\alpha]$, whose endpoints trace the triangular number $A\oplus B=(3,5,8)$. Addition of triangular numbers is therefore componentwise on the three vertices, a fact that keeps the examples in the later subsections elementary.
\end{example}

The grades produced along a computation must be combined, and the operation that combines them is a \emph{triangular norm}. A \emph{t-norm} $\otimes\colon[0,1]^2\to[0,1]$ is a commutative, associative, monotone binary operation with unit $1$; the three standard instances are the \emph{minimum} $x\wedge y$, the \emph{product} $x\cdot y$, and the \emph{{\L}ukasiewicz} t-norm $\max(0,x+y-1)$. Its dual \emph{t-conorm} (s-norm), with unit $0$, aggregates alternatives, the canonical instance being the \emph{maximum} $x\vee y$. The pair $(\wedge,\vee)$ underlies the max-min automata of Section~\ref{ssec:fuzzyauto}, while the pair $(\cdot,\vee)$ underlies their max-product variant and, as Section~\ref{ssec:weighted} shows, coincides with a weighted automaton over a particular semiring.

\begin{example}[Combining grades]\label{ex:tnorm}
Two production grades $0.8$ and $0.6$ used in one derivation combine, under the minimum t-norm, to $0.8\wedge 0.6=0.6$, and under the product t-norm to $0.8\cdot 0.6=0.48$. The product penalizes the length of a derivation, since each further step can only lower the running grade, whereas the minimum records only the weakest step; this difference, immaterial for a single step, becomes decisive when the reliability of a long derivation is at stake, as Remark~\ref{rem:indep} makes explicit.
\end{example}

\subsection{Z-numbers and their semantics}\label{ssec:znum}

\begin{figure}[t]
\centering
\includegraphics[width=\textwidth]{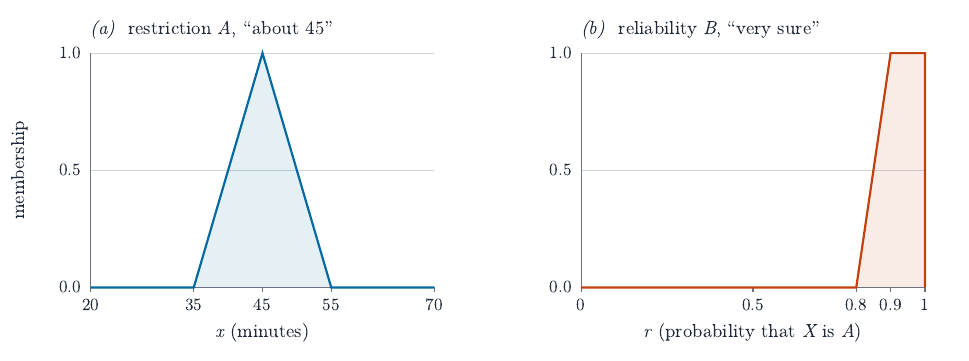}
\caption{A Z-number $Z=(A,B)$ for the valuation \emph{(about 45 minutes, very sure)}. The restriction $A$ is a triangular fuzzy number on the value axis; the reliability $B$ is a fuzzy number on $[0,1]$ constraining the probability $P(X\text{ is }A)$ of equation~\eqref{eq:pfe}.}\label{fig:znum}
\end{figure}

A \emph{Z-number}, introduced by Zadeh~\cite{zadeh2011} to represent information that is at once imprecise and only partially reliable, is an ordered pair $Z=(A,B)$ in which the first component $A$ is a fuzzy number that acts as a \emph{restriction} on the values of a real-valued uncertain variable $X$, and the second component $B$ is a fuzzy number on the unit interval $[0,1]$ that measures the \emph{reliability}, or certainty, of that restriction. Both components are, in the original setting, expressed in natural language, so that the canonical illustration is the pair $(\text{about }45\text{ min},\ \text{very sure})$ shown in Figure~\ref{fig:znum}. The precise meaning of the second component is probabilistic: writing $p_X$ for the (unknown) probability density of $X$, the quantity constrained by $B$ is the probability of the fuzzy event ``$X$ is $A$'',
\begin{equation}\label{eq:pfe}
P(X\text{ is }A)=\int \mu_A(x)\,p_X(x)\,dx,
\end{equation}
so that $B$ is a fuzzy restriction on this probability rather than on $X$ itself, and a Z-number is in this sense a \emph{second-order} object, a restriction together with a graded certainty about how probable that restriction is.

When a scalar reliability is wanted we summarize $B$ by a number $\beta(B)\in[0,1]$, which in this paper is the centroid of $B$, a conventional defuzzification that locates the constraint on $P(X\text{ is }A)$; it is a summary of that constraint and not itself an observed probability. A \emph{discrete} Z-number restricts both components to finite supports, which is the form in which the arithmetic is implemented, whereas a \emph{continuous} Z-number keeps the fuzzy numbers continuous; a related construct, the \emph{$Z^{+}$-number} $(A,R)$, pairs the restriction with an actual probability distribution $R$ rather than with a reliability about one, and serves as the computational substrate on which the arithmetic of Section~\ref{ssec:zarith} is defined.

\begin{example}[A linguistic and a data-driven Z-number]\label{ex:znum}
The valuation \emph{(about 45 minutes, very sure)} is modeled by the restriction $A=(35,45,55)$ and the trapezoidal reliability $B=(0.8,0.9,1,1)$ of Figure~\ref{fig:znum}, read as the assertion that the duration is near $45$ minutes and that this assertion is very probably correct. The second component need not be elicited; it can be derived from evidence. If a place carries a mean rating of \emph{about} $4.4$ stars computed from $n=9$ reviews, and the characteristic count of the reference class of places is estimated at $m=20$, then the evidence score $k=n/(n+m)=9/29\approx0.31$ can be used to place the reliability constraint $B$ of the Z-number $(\text{about }4.4,\ B)$ near $0.31$ rather than near one, whereas a place with $n=200$ reviews would receive $k=200/220\approx0.91$. The two second components differ in kind, one elicited and one computed from a count, and the next paragraph keeps them apart.
\end{example}

\paragraph{Three readings of reliability} Three objects that are easily conflated must be kept apart in what follows. The first is Zadeh's fuzzy constraint $B$ on the probability $P(X\text{ is }A)$ of~\eqref{eq:pfe}. The second is a scalar summary $\beta(B)$, such as the centroid, which locates that constraint but is not an observed probability and does not commute with the arithmetic of fuzzy numbers (Remark~\ref{rem:fuzzyvalued}). The third is an evidence score such as $n/(n+m)$, computed from counts, which is a modeling choice and is not, without a calibration argument, either the centroid of a constraint or a probability. The structures of Sections~3 to~5 are stated for scalar grade-reliability pairs $(a,b)\in[0,1]^{2}$ and depend on none of these readings; we call them Z-number-inspired for this reason, and we state wherever an interpretation is used which reading it requires. The probabilistic justification of the product law in Remark~\ref{rem:indep} reads the reliability as a probability, and the example of Section~6 uses an evidence score.

\subsection{The arithmetic of Z-numbers}\label{ssec:zarith}

Because the weights that decorate a grammar or an automaton will be pairs inspired by Z-numbers, the way two Z-numbers combine must be recalled. The arithmetic of Aliev et al.~\cite{aliev2015,aliev2016} computes a binary operation $Z_1\star Z_2=(A,B)$, for $\star$ among addition, subtraction, multiplication, and division, in two stages. In the first stage the restriction of the result is obtained from the restrictions of the operands alone by the extension principle~\eqref{eq:ext}, so that for addition $A=A_1\oplus A_2$ is computed exactly as in Example~\ref{ex:tri} and does not depend on $B_1$ or $B_2$. In the second stage the reliability $B$ is recovered from the probabilistic layer: among the probability distributions compatible with each operand, in the sense that they are admitted by $(A_k,B_k)$ through~\eqref{eq:pfe}, the distribution of the result is formed by convolution, and $B$ is assembled from the induced probabilities of the fuzzy event carried by $A$, an assembly that the cited works reduce to linear and small nonlinear programs.

Two structural facts matter for what follows. First, the independence of the restriction from the reliabilities means that $B$ can influence a decision only through a rule that reads it, never through the value that the restriction delivers. Second, the reliability of a sum is not an accumulation of evidence. Pooling $K$ independent measurements of one quantity raises the reliability of the pooled estimate toward one, whereas the second component of a Z-number sum constrains the probability of a different event, that the sum satisfies the restriction $A_1\oplus A_2$, and is computed from the distributions admitted by the operands, not from the number of observations behind them. \emph{Adding} Z-numbers and \emph{pooling} the evidence behind them are therefore distinct operations. Neither of them composes the steps of a derivation, and Example~\ref{ex:zsum} and Remark~\ref{rem:indep} keep the three apart.

\begin{example}[What the second component of a sum guarantees]\label{ex:zsum}
Consider two operands (\emph{about}~2, \emph{sure}) and (\emph{about}~3, \emph{sure}), with restrictions $A_1=(1,2,3)$ and $A_2=(2,3,4)$ on independent variables $X$ and $Y$. The restriction of the sum is $A_1\oplus A_2=(3,5,7)$, \emph{about} $5$, obtained vertex by vertex as in Example~\ref{ex:tri}. For its reliability, the extension principle~\eqref{eq:ext} gives, for all $x$ and $y$,
\[
\mu_{A_1\oplus A_2}(x+y)\ \ge\ \mu_{A_1}(x)\wedge\mu_{A_2}(y)\ \ge\ \mu_{A_1}(x)\,\mu_{A_2}(y),
\]
and integrating this inequality against the joint density of $(X,Y)$, which is the product of the marginal densities, gives through~\eqref{eq:pfe}
\[
P(X+Y\text{ is }A_1\oplus A_2)\ \ge\ P(X\text{ is }A_1)\cdot P(Y\text{ is }A_2).
\]
The product of the two reliabilities is therefore a \emph{lower} bound on the reliability of the sum, and the bound is not tight in general: the sum reaches the peak $5$ also at $x=y=2.5$, where $\mu_{A_1\oplus A_2}(5)=1$ while $\mu_{A_1}(2.5)\,\mu_{A_2}(2.5)=\tfrac14$. If each operand holds with probability $0.9$, the sum holds with probability at least $0.81$ and possibly more. Because the second stage of the arithmetic forms $B$ from induced probabilities of this kind, the second component need not drift downward under repeated Z-addition.
\end{example}

\begin{remark}[Reliability along a derivation]\label{rem:indep}
The reliability law used from Section~\ref{ssec:zws} on is not a consequence of Z-arithmetic; it rests on an independence assumption, which we make explicit here. A derivation asserts that every one of its steps satisfies its restriction, a conjunction and not a sum. Let the steps carry restrictions $A_1,\dots,A_n$ on variables $X_1,\dots,X_n$, and take the scalar reliability $b_i$ to be the probability $b_i=P(X_i\text{ is }A_i)$ of~\eqref{eq:pfe} itself, a reading distinct from a summary $\beta(B_i)$ of Zadeh's constraint and from an evidence score. If these restrictions hold independently, with the conjunction of fuzzy events read through the product of their membership functions, the probability that all of them hold is $b_1b_2\cdots b_n$. This product is the multiplicative reliability law of Section~\ref{ssec:zws}: a long derivation is no more reliable than its least reliable step, because it asserts more, and not because adding Z-numbers lowers their reliability.

The independence must hold between occurrences, and it fails in an important case. If a derivation uses the same estimated rule twice, and $E$ is the event that the estimate of that rule is adequate, with $P(E)=b$, then the two uses do not create two independent adequacy events, since $P(E\cap E)=P(E)=b$ and not $b^{2}$. The value $b^{2}$ belongs to a different model, in which each occurrence of the rule succeeds or fails independently with probability $b$. Rules estimated from a shared corpus are, in addition, dependent through that corpus, and the rules of one nonterminal are dependent through the multinomial distribution of its counts. We therefore distinguish a \emph{per-occurrence} reliability, for which the product is a probability under the stated independence, from \emph{epistemic confidence} in a shared parameter, for which it is not, and outside the first reading we treat the product as a declared path score rather than as a probability. The score carries a length penalty: a derivation with more steps receives a lower score even when its steps are as well attested as those of a shorter one, so that parses of different lengths are compared on unequal terms (Example~\ref{ex:zgram}), which is a limitation of the multiplicative law. Definition~\ref{def:axes} accordingly keeps the path-reliability law as a parameter.
\end{remark}

\subsection{Words, languages, grammars, and automata}\label{ssec:classical}

We recall the classical apparatus in the notation used below. An \emph{alphabet} $\Sigma$ is a finite nonempty set of \emph{symbols}; a \emph{word} (string) is a finite sequence of symbols; the \emph{empty word} is $\varepsilon$; the set of all words is the free monoid $\Sigma^{*}$ under concatenation with unit $\varepsilon$; and a \emph{language} is a subset $L\subseteq\Sigma^{*}$. A \emph{grammar} is a quadruple $G=(N,\Sigma,P,S)$ in which $N$ is a finite set of \emph{nonterminals} disjoint from $\Sigma$, $S\in N$ is the \emph{start symbol}, and $P$ is a finite set of \emph{productions} $\alpha\to\beta$ with $\alpha,\beta$ strings over $N\cup\Sigma$ and $\alpha$ containing at least one nonterminal. Writing $\Rightarrow$ for the one-step rewriting relation and $\Rightarrow^{*}$ for its reflexive-transitive closure, the \emph{language generated} by $G$ is $L(G)=\{w\in\Sigma^{*}:S\Rightarrow^{*}w\}$. The \emph{Chomsky hierarchy} grades grammars by the shape of their productions into \emph{regular} (type~3), \emph{context-free} (type~2), \emph{context-sensitive} (type~1), and \emph{unrestricted} (type~0), with the strictly increasing language families that bear the same names; a \emph{right-linear} grammar, whose productions have the forms $A\to aB$ and $A\to a$ and $A\to\varepsilon$ with $a\in\Sigma$ a single letter, is the canonical presentation of a regular grammar. For context-free grammars we count \emph{leftmost} derivations, so that each derivation tree corresponds to exactly one derivation and the different rewriting orders of one tree are not counted separately; for general grammars a derivation is a sequence of one-step rewritings.

A \emph{nondeterministic finite automaton} is a quintuple $M=(Q,\Sigma,\Delta,q_0,F)$ with $Q$ a finite set of \emph{states}, $\Delta\colon Q\times\Sigma\to 2^{Q}$ a \emph{transition function}, $q_0\in Q$ the \emph{initial} state, and $F\subseteq Q$ the \emph{accepting} states; $M$ \emph{accepts} $w$ when some path labeled by $w$ leads from $q_0$ into $F$, and the \emph{recognized} language is the set of accepted words. The classical equivalence that organizes the later generalization is that the right-linear grammars and the finite automata describe exactly the same family, the regular languages, each convertible into the other by reading nonterminals as states and productions as transitions.

\begin{example}[A context-free and a regular description]\label{ex:classical}
The grammar with productions $S\to aSb\mid\varepsilon$ generates the nonregular context-free language $\{a^{n}b^{n}:n\ge0\}$, since each use of the first production adds a matching pair and the derivation $S\Rightarrow aSb\Rightarrow aaSbb\Rightarrow aabb$ produces $a^{2}b^{2}$. By contrast, the right-linear grammar $S\to aT$, $T\to aT\mid b$ generates the regular language $a^{+}b$ of one or more $a$'s followed by a single $b$. Reading $S$ as an initial state $q_0$, $T$ as a state $q_1$, and the terminating production $T\to b$ as a transition into an accepting state $q_2$ turns it into the automaton that reads $a$ from $q_0$ to $q_1$, loops on $a$ at $q_1$, and reads $b$ from $q_1$ to $q_2$. This automaton recognizes the same language, and it is the automaton drawn, with weights added, in Figure~\ref{fig:wauto}.
\end{example}

\subsection{Fuzzy grammars and fuzzy automata}\label{ssec:fuzzyauto}

The first graded generalization, due to Lee and Zadeh~\cite{lee1969}, attaches to each production a degree of membership. A \emph{fuzzy grammar} is a system $G=(N,\Sigma,P,S,\varphi)$ in which $(N,\Sigma,P,S)$ is a grammar and $\varphi\colon P\to[0,1]$ assigns to each production a \emph{grade}; the grade of a derivation $S\Rightarrow^{*}w$ is the t-norm (classically the minimum, or alternatively the product) of the grades of the productions it uses, and the \emph{fuzzy language} generated is the membership function
\begin{equation}\label{eq:fuzzylang}
\mu_{L(G)}(w)=\sup\bigl\{\,\varphi(\text{derivation}):\text{derivation of }w\,\bigr\},
\end{equation}
the supremum being taken over all derivations of $w$, so that $\mu_{L(G)}\colon\Sigma^{*}\to[0,1]$ is a genuine fuzzy subset of $\Sigma^{*}$. The automaton-side counterpart is a \emph{fuzzy finite automaton}, in which the transition function is replaced by a map $\Delta\colon Q\times\Sigma\times Q\to[0,1]$ grading each transition, and the membership of a word is the maximum over paths of the minimum (or the product) of the transition grades along the path, the construction known as a \emph{max-min} (respectively \emph{max-product}) automaton. Replacing the unit interval by a lattice-ordered monoid, as Li and Pedrycz~\cite{li2005} do for fuzzy finite automata, extends the same idea beyond $[0,1]$ and prepares the move to arbitrary weight structures made in the next subsection.

\begin{example}[A fuzzy right-linear grammar]\label{ex:fuzzygram}
Grade the productions of the right-linear grammar of Example~\ref{ex:classical} as $S\to aT\ (0.9)$, $T\to aT\ (0.5)$, $T\to b\ (0.7)$ and read the grade of a derivation under the product t-norm. The word $aab$ has a unique leftmost derivation $S\Rightarrow aT\Rightarrow aaT\Rightarrow aab$, using the productions graded $0.9$, $0.5$, $0.7$, so its grade is $0.9\cdot0.5\cdot0.7=0.315$, and since no other derivation yields $aab$ the supremum in~\eqref{eq:fuzzylang} gives $\mu_{L(G)}(aab)=0.315$. The word $ab$ has the single derivation $S\Rightarrow aT\Rightarrow ab$ of grade $0.9\cdot0.7=0.63$, so the shorter word is the more strongly generated, which is the qualitative signature of the product reading already noted in Example~\ref{ex:tnorm}.
\end{example}

\subsection{Weighted automata and their weight structures}\label{ssec:weighted}

The fuzzy automata of the previous subsection are a special case of a construction in which the grades are drawn from an algebraic structure chosen to fit the quantity being computed. A \emph{semiring} is a tuple $(\Sem,\oplus,\otimes,\mathbf{0},\mathbf{1})$ in which $(\Sem,\oplus,\mathbf{0})$ is a commutative monoid, $(\Sem,\otimes,\mathbf{1})$ is a monoid, multiplication distributes over addition on both sides, and $\mathbf{0}$ annihilates under multiplication. The canonical instances, collected in Table~\ref{tab:semirings}, specialize the same machinery to Boolean recognition, to probability, to shortest paths, and to fuzzy membership, the last of these being exactly the max-product reading of Section~\ref{ssec:fuzzyauto}. A \emph{weighted finite automaton} over $\Sem$ assigns to each transition a weight in $\Sem$, and optionally initial and final weights; the weight of a path is the product $\otimes$ of its transition weights, and the \emph{behavior} of the automaton is the function
\begin{equation}\label{eq:behavior}
\sem{\mathcal{A}}(w)={\textstyle\bigoplus_{\pi\ \text{a path for }w}}\ \operatorname{wt}(\pi)\ \in\ \Sem,
\end{equation}
the sum $\oplus$ being taken over all paths labeled by $w$, so that the behavior is a formal power series $\sem{\mathcal{A}}\colon\Sigma^{*}\to\Sem$. The organizing result of the theory, the \emph{Kleene--Schützenberger theorem}, states that the series realized by weighted automata over a semiring are exactly the \emph{rational} series, those built from the letters by sum, product, and iteration, generalizing Kleene's theorem for the Boolean case.

\begin{table}[t]
\centering\small
\caption{Common weight structures and the quantity each makes a weighted automaton compute.}\label{tab:semirings}
\begin{tabular}{@{}lllll@{}}
\toprule
Semiring & Domain & $\oplus$ & $\otimes$ & Behavior computes\\
\midrule
Boolean & $\{0,1\}$ & $\vee$ & $\wedge$ & acceptance (a language)\\
Probability & $[0,\infty)$ & $+$ & $\times$ & total path mass (a word probability\\
 & & & & \quad under stochastic normalization)\\
Viterbi & $[0,1]$ & $\max$ & $\times$ & best-path grade (max-product fuzzy)\\
Tropical (min-plus) & $\Real\cup\{+\infty\}$ & $\min$ & $+$ & least-cost (shortest) path\\
Fuzzy (max-min) & $[0,1]$ & $\max$ & $\min$ & best-path membership\\
\bottomrule
\end{tabular}
\end{table}

\begin{figure}[t]
\centering
\includegraphics[width=0.968\textwidth]{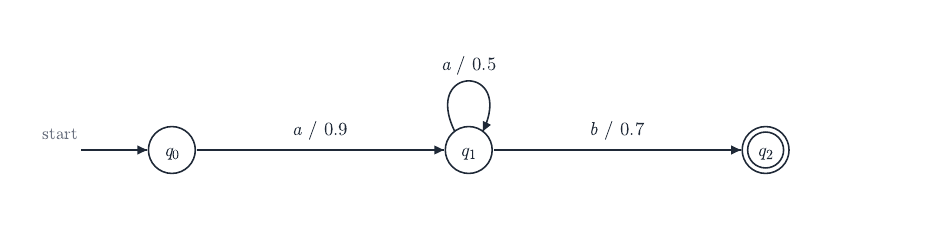}
\caption{A weighted automaton on a regular skeleton (Example~\ref{ex:wauto}). Each transition carries a label and a weight; the behavior aggregates path weights by the operations of the chosen weight structure, here the Viterbi semiring.}\label{fig:wauto}
\end{figure}

\begin{example}[One automaton, several weight structures]\label{ex:wauto}
Figure~\ref{fig:wauto} shows a weighted automaton on the regular skeleton of Example~\ref{ex:classical}, with the transition $q_0\xrightarrow{a}q_1$ weighted $0.9$, the self-loop $q_1\xrightarrow{a}q_1$ weighted $0.5$, and $q_1\xrightarrow{b}q_2$ weighted $0.7$, with initial weight $1$ on $q_0$, final weight $1$ on $q_2$, and all other initial and final weights $0$. Over the Viterbi semiring $([0,1],\max,\times)$ the word $aab$ has the single accepting path $q_0\,q_1\,q_1\,q_2$ of weight $0.9\times0.5\times0.7=0.315$, so $\sem{\mathcal{A}}(aab)=0.315$, which reproduces the grade computed from the fuzzy grammar of Example~\ref{ex:fuzzygram}. Under the correspondence of Example~\ref{ex:classical}, each production of that grammar becomes the transition of the same grade, so the fuzzy grammar is this weighted automaton in disguise. Reinterpreting the very same weights over the tropical semiring $(\Real\cup\{+\infty\},\min,+)$, the behavior of $aab$ becomes $0.9+0.5+0.7=2.1$, now read as the least accumulated cost of reading the word, which shows how the choice of weight structure, and not the automaton, fixes the meaning of the computation.
\end{example}

The requirement that the weights form a full semiring is stronger than the definition of the behavior~\eqref{eq:behavior} actually needs. When distributivity is dropped and only the two monoid structures are retained, the weights form a \emph{strong bimonoid}, over which weighted automata remain well defined although the run semantics and the initial-algebra semantics may diverge~\cite{dsv2010}. A Chomsky--Schützenberger-style characterization has been established over weight structures still more general than semirings, encompassing valuation monoids and averaging operations~\cite{dv2014}. Strong bimonoids are the general setting in which several natural ways of combining Z-weights live, and several of the structures classified in Table~\ref{tab:class} are strong bimonoids but not semirings. The structure on which the paper rests is, however, a commutative semiring (Proposition~\ref{prop:semiring}). The full theory of weighted automata over semirings therefore applies to it, and the weaker rungs of this hierarchy serve to locate its alternatives, not to admit it.

\subsection{Related graded devices}\label{ssec:cw}

Zadeh's program of \emph{computing with words} produced one of the formalisms related to the one developed here, and the distinction between the two must be stated at the outset. In the model of Ying~\cite{ying2002}, a fuzzy automaton is fed not strings of symbols but strings of \emph{fuzzy subsets} of the alphabet, that is, words in the sense of linguistic terms, and an extension principle carries computing with values up to computing with words. Wang and Qiu~\cite{wang2003} carry the same approach from fuzzy automata to Turing machines, and Cao, Ying, and Chen~\cite{cao2007} study the retraction and the generalized extension of computing with words. In every case the object that is generalized is the \emph{input}, whose symbols become words while the transition weight remains a scalar grade drawn from $[0,1]$. The direction pursued below is orthogonal: the input alphabet stays classical, and it is the \emph{weight} that is generalized, from a scalar grade to a scalar grade-reliability pair inspired by a Z-number $(A,B)$, which carries both a value and its reliability. The next section composes such weights along a derivation and selects among competing derivations by their \emph{credible grade}, the grade discounted by its reliability. Keeping these two axes of generalization apart, the symbol axis of computing with words and the weight axis of the present work, is essential to locating the contribution against the literature.

A second group of relatives carries more than one quantity along a path. Products of semirings aggregate each component independently; lexicographic semirings select by a first component and use a second to break ties~\cite{roark2011}; selective semirings return one operand of every sum and correspond to total orders compatible with the product~\cite{gondran2008}; and the derivation semirings of semiring parsing carry the best derivation along with its score~\cite{goodman1999}. Intuitionistic fuzzy automata pair a degree of membership with a degree of nonmembership~\cite{jun2005,zhang2009}. The structure of Section~3 belongs to this group algebraically, being a selective semiring with a lexicographic description, and Section~\ref{ssec:compare} compares it with these devices in detail. What distinguishes it is the role of its second component, which qualifies the first rather than describing the word.

\FloatBarrier
\section{Z-grammars and Z-automata}

The preliminaries assembled the three ingredients of the construction, namely a Z-number-inspired pair as a two-component weight, a grammar or an automaton as its carrier, and the weighted-automata apparatus of Section~\ref{ssec:weighted} as the algebraic frame. We now bring them together. The guiding principle is that the single graded quantity of a fuzzy grammar, the production grade of Section~\ref{ssec:fuzzyauto}, is replaced by a \emph{Z-weight}, an ordered pair that records both a \emph{grade}, how strongly the production or transition fires, and a \emph{reliability}, how much that grade is to be trusted. The two components are propagated along a derivation or a run by operations chosen so that grades compose as degrees and reliabilities compose as trust, while the competing derivations of a word are weighed by their grade and their reliability together rather than by either alone. A Z-grammar and a Z-automaton are then, respectively, a weighted grammar and a weighted automaton over the resulting weight structure. The behavior of each is a formal power series whose values are scalar grade-reliability pairs, assigning to every word not a single membership degree but a membership degree together with its reliability; fuzzy-valued weights are discussed, and left open, in Remark~\ref{rem:fuzzyvalued}.

\subsection{The Z-weight structure}\label{ssec:zws}

So that the definitions and the examples remain transparent, we work throughout this section with the scalar summaries of the two components, writing a Z-weight as a pair $z=(a,b)$ with $a$ the grade and $b$ a scalar reliability, which in an application may be a summary $\beta(B)$ or an evidence score (Section~\ref{ssec:znum}). The fully fuzzy-valued version, in which $a$ and $b$ are themselves fuzzy numbers, is not treated here (Remark~\ref{rem:fuzzyvalued} and Section~7). The reading of a weight makes a quantity decisive only to the extent that it is both high and trusted, so the quantity that orders competing derivations is the \emph{credible grade} $\cred(a,b)=a\,b$, the grade discounted by its reliability. We write $\grade(a,b)=a$ and $\rel(a,b)=b$ for the two projections, so that $\cred(z)=\grade(z)\,\rel(z)$. The credible grade should not be confused with the credibility measure of fuzzy-variable theory~\cite{liu2002}, a self-dual set function defined as the average of possibility and necessity; the credible grade is a number attached to a single weight, its grade discounted by its reliability.

\begin{definition}[Z-weight structure]\label{def:zws}
The \emph{Z-weight structure} is $\Zset=(Z,\oplus,\otimes,\mathbf{0},\mathbf{1})$ on the \emph{canonical carrier} $Z=\Zzz=\{\mathbf{0}\}\cup(0,1]^{2}$, in which $\mathbf{0}$ is a distinguished element, written $(0,0)$, and $\mathbf{1}=(1,1)$. The product of two elements of $(0,1]^{2}$ is
\begin{equation}\label{eq:zotimes}
(a_1,b_1)\otimes(a_2,b_2)=(a_1a_2,\ b_1b_2),
\end{equation}
which again lies in $(0,1]^{2}$, and $\mathbf{0}\otimes z=z\otimes\mathbf{0}=\mathbf{0}$ for every $z\in Z$. The carrier is totally ordered by letting $\mathbf{0}\prec z$ for every $z\neq\mathbf{0}$ and, for $z_1=(a_1,b_1)$ and $z_2=(a_2,b_2)$ in $(0,1]^{2}$, $z_1\prec z_2$ when $a_1b_1<a_2b_2$, or when $a_1b_1=a_2b_2$ and $a_1<a_2$; two elements of $(0,1]^{2}$ with equal credible grade $\cred(a,b)=a\,b$ and equal grade are equal, so no further tie-break is needed. The sum of a finite family is its greatest element in this order,
\begin{equation}\label{eq:zoplus}
{\textstyle\bigoplus_{i\in I}}(a_i,b_i)=(a_{i^{*}},b_{i^{*}}),\qquad (a_{i^{*}},b_{i^{*}})=\max\nolimits_{\preceq}\{(a_i,b_i):i\in I\},
\end{equation}
and the empty sum is $\mathbf{0}$, so that in the binary case $(a_1,b_1)\oplus(a_2,b_2)$ reports the operand of larger credible grade $a\,b$, ties broken toward the larger grade.
\end{definition}

The canonical carrier may equivalently be read as the quotient of $[0,1]^{2}$ that identifies every weight of grade $0$ \emph{or} reliability $0$ with $\mathbf{0}$. Under selection by the credible grade this identification is compatible with both operations (Proposition~\ref{prop:quotcong}), so the two descriptions give the same structure.

The reading of the two operations is the one announced above. The product~\eqref{eq:zotimes} composes two steps taken in sequence. It multiplies their grades as independent degrees of firing, and it multiplies their reliabilities because, under the independence assumption made explicit in Section~\ref{ssec:zarith}, the probability that the restrictions of both steps hold is the product of the probabilities that each holds; a long derivation is therefore both weakly graded and weakly reliable. The sum~\eqref{eq:zoplus} weighs the competing derivations of one word by selecting the \emph{most credible}, the derivation whose grade best survives its reliability, and it reports the grade and the reliability of that derivation together. The aggregated value thus answers at once how strongly and how reliably the word is obtained, and a high grade on a poorly trusted derivation cannot override a slightly lower grade that is far better trusted. The credible grade is the quantity on which the structure is organized. The map $\cred$ carries $\Zset$ onto the Viterbi semiring $([0,1],\vee,\cdot)$ of Section~\ref{ssec:weighted}, so that a Z-device sits directly above an ordinary max-product device, now on the discounted grade rather than on the raw grade, with the two components carried along the computation that the credible grade performs.

The structure belongs to a familiar class, and it is worth placing it there before stating its laws. A semiring is \emph{selective} when its sum always returns one of its operands~\cite{gondran2008}. Commutative selective semirings are exactly the structures obtained from a commutative monoid $(M,\otimes,\mathbf{1})$ with an absorbing element $\mathbf{0}$ and a total order $\preceq$ on $M$, with least element $\mathbf{0}$, that is compatible with the product in the sense that $x\preceq y$ implies $x\otimes z\preceq y\otimes z$, by letting $\oplus$ return the larger operand: compatibility is exactly the distributivity of this maximum over $\otimes$, and conversely the sum of a selective semiring defines such an order by setting $x\preceq y$ when $x\oplus y=y$. The Viterbi and the tropical semirings are the familiar instances, and the closest in spirit is the Viterbi-derivation semiring of semiring parsing~\cite{goodman1999}, which carries the best derivation along with its score. Proposition~\ref{prop:semiring} is therefore not new in its form. Its content lies in the choice of the order, the credible grade refined by the grade, and in the carrier $\Zzz$ that makes this order compatible with the product. On the full square the order is not compatible: $(0.9,0.1)$ precedes $(0.5,0.5)$ by credible grade, yet after multiplication by $(1,0)$ the images $(0.9,0)$ and $(0.5,0)$, both of credible grade $0$, are ordered the other way by the grade. Section~4 shows which of the alternative orders and carriers retain the laws and which do not.

The order of Definition~\ref{def:zws} has a transparent description that places $\Zset$ among known structures. On $\Zzz$ the map $\eta(a,b)=(a\,b,\ a)$, with $\eta(\mathbf{0})=(0,0)$, is injective, since a weight other than $\mathbf{0}$ has $a>0$ and its reliability is recovered as $b=(a\,b)/a$. Its image consists of $(0,0)$ and the pairs $(u,v)$ with $0<u\le v\le1$, a set closed under the componentwise product. Let the \emph{lexicographic product} of two Viterbi semirings carry the componentwise product and the sum that returns the lexicographically larger of two pairs. On the pairs with both coordinates positive, together with $(0,0)$, it is a commutative semiring, because multiplication by such a pair is strictly increasing in each coordinate. The map $\eta$ sends $\otimes$ to the componentwise product, and it sends $\oplus$ to the lexicographic maximum, because the order of Definition~\ref{def:zws} compares the credible grade first and the grade second, and $(a\,b,a)$ determines $b$. Hence $\eta$ identifies $\Zset$ with a subsemiring of this lexicographic product. Neither this embedding nor the semiring laws that follow from it are new as general theorems; we record them because they fix the structure precisely. The Z-weight structure is thus a lexicographic semiring on the pair (credible grade, grade), and not on the pair (grade, reliability): the reliability enters the primary comparison as a discount on the grade and is never consulted on its own.

\begin{proposition}[$\Zset$ is a commutative semiring]\label{prop:semiring}
The Z-weight structure $\Zset$ of Definition~\ref{def:zws} is a commutative semiring, and the credible-grade map $\cred\colon\Zset\to([0,1],\vee,\cdot)$, $\cred(a,b)=a\,b$, is a semiring homomorphism onto the Viterbi semiring.
\end{proposition}

\begin{proof}
We verify the laws directly; the core of the argument, the distributivity step, is the compatibility of the order with the product. That $(Z,\otimes,\mathbf{1})$ is a commutative monoid is immediate from~\eqref{eq:zotimes}. The sum~\eqref{eq:zoplus} selects, from the unordered family, the element greatest in the total order of Definition~\ref{def:zws}; selection by a fixed total order is associative, commutative, and idempotent, so $(Z,\oplus,\mathbf{0})$ is a commutative idempotent monoid, in which $\mathbf{0}$ is neutral because it is the least element. By definition $\mathbf{0}$ is absorbing for $\otimes$. For distributivity, the credible grade of a product is $\cred\bigl((a_1,b_1)\otimes(a_i,b_i)\bigr)=(a_1b_1)(a_ib_i)$, so multiplying a family by a weight $(a_1,b_1)$ of positive credible grade scales every credible grade by the same positive constant $a_1b_1$ while multiplying every grade by $a_1>0$ and every reliability by $b_1>0$; since multiplication by positive constants preserves both the credible-grade order and the tie-break by grade, the maximizer that~\eqref{eq:zoplus} selects is unchanged, so the two sides of the distributive law select the same operand and scale it identically. A factor $\mathbf{0}$ reduces both sides to $\mathbf{0}$, and commutativity of $\otimes$ gives the law on both sides. Finally $\cred$ sends $\otimes$ to the product, since $\cred(a_1a_2,b_1b_2)=a_1a_2b_1b_2=\cred(a_1,b_1)\cred(a_2,b_2)$, and $\oplus$ to the maximum, since the selected operand is the one of greatest credible grade, and it preserves the units, so it is a homomorphism onto the Viterbi semiring. The distributive identity was confirmed by exact rational computation over the grid $\{0,\tfrac14,\tfrac12,\tfrac34,1\}^{2}$; the classification of this structure among representative alternatives is the subject of Section~4.
\end{proof}

A first consequence is that the apparatus of Section~\ref{ssec:weighted} applies without modification. By the Kleene--Schützenberger theorem the Z-languages realized by the Z-automata of Section~\ref{ssec:zauto} are exactly the $\Zset$-rational series (Theorem~\ref{thm:ks}), and the closure constructions for weighted automata over a commutative semiring, such as the Hadamard product, apply to Z-automata as stated. Decision questions are not settled by the semiring property alone; they are treated in Section~5.

\begin{remark}[The plain best-grade variant]\label{rem:bestgrade}
Selecting by the raw grade $a$ rather than by the credible grade $a\,b$ gives the simpler \emph{best-grade} structure, the select-by-grade scheme of Section~4, which is also a commutative semiring, on the quotient $\Zz$ that identifies only the grade-$0$ weights with $\mathbf{0}$. The two structures differ exactly where grade and reliability pull apart. By choosing on grade alone, the best-grade variant reproduces the defect that Example~\ref{ex:ambig} exhibits, a high grade on thin evidence overriding a slightly lower grade that is far better trusted. This is why the credible grade of Definition~\ref{def:zws} is taken as canonical here. The precise carriers on which each variant is a semiring, and the structures under which the semiring property fails altogether, are established in Section~4.
\end{remark}

\begin{remark}[Fuzzy-valued Z-weights]\label{rem:fuzzyvalued}
The scalar structure studied here is recovered exactly from singleton fuzzy-valued weights: when every grade and every reliability is a singleton fuzzy number, the extension principle~\eqref{eq:ext} reduces the product of two of them to the singleton of the product, and the construction is that of Definition~\ref{def:zws}. For general fuzzy numbers, centroid defuzzification need not commute with multiplication, and so supplies no order on fuzzy credible grades that is known to be compatible with their product. Let $A=(0,0,1)$ be the left-shoulder triangular fuzzy number with membership $1-x$ on $[0,1]$, so that $\acut{\alpha}A=[0,1-\alpha]$, and let $C$ be the extension-principle product of two copies of $A$, treated as two fuzzy arguments. Then $\acut{\alpha}C=[0,(1-\alpha)^{2}]$ and $\mu_C(y)=1-\sqrt{y}$ on $[0,1]$, so that the centroid of $A$ is $1/3$, while the centroid of $C$ is $(1/2-2/5)/(1-2/3)=3/10\neq1/9$. We therefore make no claim that a general fuzzy-valued extension projects homomorphically onto this scalar semiring, and the examples below are statements about scalar weights only. A fuzzy-valued theory would need explicit operations on fuzzy grades and reliabilities together with an order on fuzzy credible grades that is compatible with their product; it is listed among the open problems of Section~7.
\end{remark}

\subsection{Z-grammars}\label{ssec:zgram}

\begin{definition}[Z-grammar]\label{def:zgram}
A \emph{Z-grammar} is a system $G=(N,\Sigma,P,S,\zeta)$ in which $(N,\Sigma,P,S)$ is a grammar in the sense of Section~\ref{ssec:classical} and $\zeta\colon P\to Z$ assigns a Z-weight to each production. The \emph{weight} of a derivation $D\colon S\Rightarrow^{*}w$ that applies the productions $p_1,\dots,p_k$ in order, a leftmost derivation when $G$ is context-free (Section~\ref{ssec:classical}), is the Z-product of their weights,
\begin{equation}\label{eq:derwt}
\operatorname{wt}(D)=\zeta(p_1)\otimes\cdots\otimes\zeta(p_k),
\end{equation}
and the \emph{Z-language} generated by $G$ is the map $\Zlang_G\colon\Sigma^{*}\to Z$ given by
\begin{equation}\label{eq:zlang}
\Zlang_G(w)={\textstyle\bigoplus_{\,D\colon S\Rightarrow^{*}w}}\operatorname{wt}(D),
\end{equation}
the Z-sum over all derivations of $w$, of which there may be infinitely many (Remark~\ref{rem:infsum}), with the empty sum equal to $\mathbf{0}=(0,0)$, so that a word with no derivation receives the Z-weight $(0,0)$.
\end{definition}

\begin{remark}[Sums over infinitely many derivations]\label{rem:infsum}
A grammar may give a word infinitely many derivations, for instance through a cycle of unit productions, so the Z-sum in~\eqref{eq:zlang} must be defined for infinite families. We take it to be the greatest element of the family in the total order by which~\eqref{eq:zoplus} selects, and this element exists. A derivation that uses a production of weight $\mathbf{0}$ has weight $\mathbf{0}$, and since products of elements of $(0,1]^{2}$ stay in $(0,1]^{2}$, every other derivation has a nonzero weight that is a product of elements of the finite set $W\subseteq Z\setminus\{\mathbf{0}\}$ of nonzero production weights of $G$. A weight of credible grade $1$ equals $\mathbf{1}$, since $a\,b=1$ forces $a=b=1$. Let $\lambda<1$ be the largest credible grade of a weight in $W\setminus\{\mathbf{1}\}$, with $\lambda=0$ when there is none. Since the credible grade is multiplicative, a product of weights in $W$ whose credible grade is at least $\theta>0$ has at most $\ell$ factors different from $\mathbf{1}$, where $\ell=\lfloor\log\theta/\log\lambda\rfloor$ when $\lambda>0$ and $\ell=0$ when $\lambda=0$; such products therefore take only finitely many values. If every derivation of $w$ has weight $\mathbf{0}$, the sum is $\mathbf{0}$. Otherwise some derivation has a positive credible grade $\theta$, the derivation weights of credible grade at least $\theta$ form a finite nonempty set of values, and the greatest of these is the greatest element of the whole family. Infinitely many derivations may share this greatest weight, for example when they differ only by factors equal to $\mathbf{1}$, and the sum then returns that weight once, as the idempotence of $\oplus$ requires. For a finite family the definition agrees with~\eqref{eq:zoplus}.
\end{remark}

A Z-language records, for each word, the grade and the reliability of its most credible derivation, and two crisp languages are recovered from it by thresholding. For a credible-grade threshold $\theta\in(0,1]$ the \emph{$\theta$-level language} is $L_\theta(G)=\{w:\cred(\Zlang_G(w))\ge\theta\}$, the ordinary fuzzy-language cut inherited through the credible-grade projection, and for a reliability floor $\rho\in(0,1]$ the \emph{reliability-filtered language} is $L_{\theta,\rho}(G)=\{w:\cred(\Zlang_G(w))\ge\theta\ \text{and}\ \rel(\Zlang_G(w))\ge\rho\}$, which admits a word only when its most credible derivation is both strong and reliably so.

\begin{example}[A right-linear Z-grammar]\label{ex:zgram}
Let $G$ have the right-linear productions
\[
S\to aT\ :(0.9,0.9),\qquad T\to aT\ :(0.5,0.6),\qquad T\to b\ :(0.7,0.8),
\]
over $\Sigma=\{a,b\}$, generating the language $a^{+}b$ of Section~\ref{ssec:classical} with every word weighted. The word $aab$ has the single derivation $S\Rightarrow aT\Rightarrow aaT\Rightarrow aab$, whose weight by~\eqref{eq:derwt} is $(0.9,0.9)\otimes(0.5,0.6)\otimes(0.7,0.8)=(0.315,0.432)$, so $\Zlang_G(aab)=(0.315,0.432)$, with credible grade $\cred=0.315\cdot0.432=0.13608\approx0.136$. The word $ab$ has the single derivation $S\Rightarrow aT\Rightarrow ab$ of weight $(0.9,0.9)\otimes(0.7,0.8)=(0.63,0.72)$, so $\Zlang_G(ab)=(0.63,0.72)$ with credible grade $0.4536\approx0.454$, the shorter word being both more strongly generated and more reliable, as the multiplicative composition of~\eqref{eq:zotimes} dictates. Part of the difference is the length penalty of Remark~\ref{rem:indep}, since $aab$ uses one more step whatever the attestation of that step. A word such as $b$, which the grammar cannot derive, receives $\Zlang_G(b)=(0,0)$. At the credible-grade threshold $\theta=0.4$ the level language $L_{0.4}(G)$ contains $ab$ but not $aab$.
\end{example}

\begin{figure}[t]
\centering
\includegraphics[width=\textwidth]{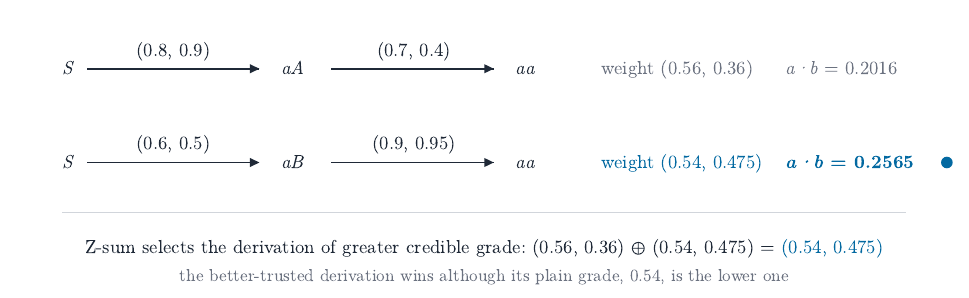}
\caption{The two derivations of $aa$ in the Z-grammar of Example~\ref{ex:ambig}, each weighted by the Z-product of its productions and aggregated by the Z-sum, which selects the derivation of greater credible grade $a\,b$. The better-trusted derivation is selected although its plain grade is the lower one.}\label{fig:zagg}
\end{figure}

\begin{example}[Ambiguity and the Z-sum]\label{ex:ambig}
Let $G'$ have the productions $S\to aA\ :(0.8,0.9)$, $S\to aB\ :(0.6,0.5)$, $A\to a\ :(0.7,0.4)$, and $B\to a\ :(0.9,0.95)$, so that the word $aa$ has the two derivations shown in Figure~\ref{fig:zagg}. The first, through $A$, has weight $(0.8,0.9)\otimes(0.7,0.4)=(0.56,0.36)$ and credible grade $0.2016$; the second, through $B$, has weight $(0.6,0.5)\otimes(0.9,0.95)=(0.54,0.475)$ and credible grade $0.2565$. Aggregating by the Z-sum~\eqref{eq:zoplus}, the more credible derivation wins, $0.2565>0.2016$, so $\Zlang_{G'}(aa)=(0.54,0.475)$, even though its plain grade $0.54$ is the lower of the two. The best-grade variant of Remark~\ref{rem:bestgrade} would instead select the first derivation on its grade $0.56$ and report the less reliable value $(0.56,0.36)$; the two readings disagree exactly where grade and reliability pull apart, and the credible grade resolves the disagreement in favor of the better-trusted derivation, which is the behavior that makes the reliability component worth carrying.
\end{example}

\subsection{Z-automata}\label{ssec:zauto}

\begin{definition}[Z-automaton]\label{def:zauto}
A \emph{Z-automaton} is a system $\mathcal{A}=(Q,\Sigma,\Delta,\mathrm{in},\zeta,\mathrm{out})$ in which $Q$ is a finite set of states, $\Delta\colon Q\times\Sigma\to2^{Q}$ is a transition function as in Section~\ref{ssec:classical}, $\zeta$ assigns a nonzero Z-weight $\zeta(q,x,q')\in Z\setminus\{\mathbf{0}\}$ to each transition $q'\in\Delta(q,x)$, a transition of weight $\mathbf{0}$ being simply omitted from $\Delta$, and $\mathrm{in},\mathrm{out}\colon Q\to Z$ are the \emph{initial} and the \emph{final} weight functions. The \emph{weight} of a path $\pi=p_0\xrightarrow{x_1}p_1\cdots\xrightarrow{x_n}p_n$ is the Z-product $\operatorname{wt}(\pi)=\bigotimes_{j=1}^{n}\zeta(p_{j-1},x_j,p_j)$, equal to $\mathbf{1}$ when $n=0$, and the \emph{behavior} of $\mathcal{A}$ is the map $\beh{\mathcal{A}}\colon\Sigma^{*}\to Z$,
\begin{equation}\label{eq:zbeh}
\beh{\mathcal{A}}(w)={\textstyle\bigoplus_{\,\pi\ \text{labeled by }w}}\mathrm{in}(p_0)\otimes\operatorname{wt}(\pi)\otimes\mathrm{out}(p_n),
\end{equation}
the Z-sum over all paths labeled by $w$, with the empty sum equal to $\mathbf{0}$. A path is \emph{accepting} when its complete product $\mathrm{in}(p_0)\otimes\operatorname{wt}(\pi)\otimes\mathrm{out}(p_n)$ is nonzero; since the transition weights are nonzero and products of elements of $(0,1]^{2}$ are nonzero, this holds exactly when $\mathrm{in}(p_0)\neq\mathbf{0}$ and $\mathrm{out}(p_n)\neq\mathbf{0}$, and only accepting paths contribute to~\eqref{eq:zbeh}. The automaton is \emph{deterministic} when at most one state has a nonzero initial weight and $|\Delta(q,x)|\le1$ for every state $q$ and letter $x$, \emph{unambiguous} when every word labels at most one accepting path, and \emph{trim} when every state lies on some accepting path.
\end{definition}

The initial and final weights carry the constant term: the behavior on the empty word is $\bigoplus_{q\in Q}\mathrm{in}(q)\otimes\mathrm{out}(q)$, which may be any element of $Z$, as the rational series of Section~5 require. When an automaton is presented, as in Section~\ref{ssec:classical}, by an initial state $q_0$ and a set $F$ of accepting states, we take $\mathrm{in}(q_0)=\mathbf{1}$, $\mathrm{out}(q)=\mathbf{1}$ for $q\in F$, and all other initial and final weights equal to $\mathbf{0}$, so that the behavior is carried entirely by the transition weights. A word of length $n$ labels only finitely many paths, so the sum~\eqref{eq:zbeh} is always finite.

\begin{figure}[t]
\centering
\includegraphics[width=0.968\textwidth]{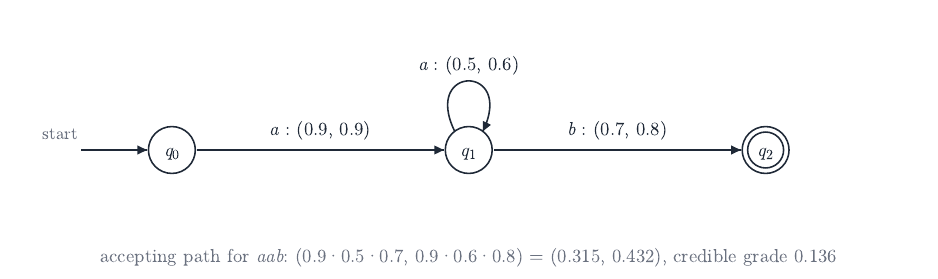}
\caption{A Z-automaton (Example~\ref{ex:zauto}). Each transition carries a Z-weight $(\text{grade},\text{reliability})$; the behavior aggregates path weights by the Z-sum, returning a grade of acceptance together with its reliability.}\label{fig:zauto}
\end{figure}

\begin{example}[The running Z-automaton]\label{ex:zauto}
Figure~\ref{fig:zauto} shows the Z-automaton on the skeleton of Section~\ref{ssec:classical} with $\zeta(q_0,a,q_1)=(0.9,0.9)$, $\zeta(q_1,a,q_1)=(0.5,0.6)$, and $\zeta(q_1,b,q_2)=(0.7,0.8)$, the state $q_0$ initial and the state $q_2$ accepting, so that $\mathrm{in}(q_0)=\mathrm{out}(q_2)=\mathbf{1}$ and all other initial and final weights are $\mathbf{0}$. The word $aab$ has the single accepting path $q_0q_1q_1q_2$, whose weight is $(0.9,0.9)\otimes(0.5,0.6)\otimes(0.7,0.8)=(0.315,0.432)$, so $\beh{\mathcal{A}}(aab)=(0.315,0.432)$, a grade of acceptance of $0.315$ together with its reliability $0.432$; the word $ab$ has the path $q_0q_1q_2$ of weight $(0.63,0.72)$. These are exactly the Z-weights that the right-linear Z-grammar of Example~\ref{ex:zgram} assigns to the same words, which is no coincidence.
\end{example}

\begin{proposition}[Equivalence of right-linear Z-grammars and Z-automata]\label{prop:equiv}
For every right-linear Z-grammar $G$ there is a Z-automaton $\mathcal{A}$ with $\beh{\mathcal{A}}=\Zlang_G$, and for every Z-automaton there is a right-linear Z-grammar generating its behavior.
\end{proposition}

\begin{proof}
Let $G=(N,\Sigma,P,S,\zeta)$ be right-linear, so that its productions have the forms $A\to xB$, $A\to x$, and $A\to\varepsilon$ of Section~\ref{ssec:classical}, with $x\in\Sigma$ a single letter. Take the states $Q=N\cup\{f\}$ with a new state $f$, and set $\mathrm{in}(S)=\mathbf{1}$ and $\mathrm{in}(q)=\mathbf{0}$ for $q\neq S$. Read each production $A\to xB$ as the transition $A\xrightarrow{x}B$ and each production $A\to x$ as the transition $A\xrightarrow{x}f$, carrying the weight of the production; since $P$ is a set, distinct productions give distinct transitions. Set $\mathrm{out}(f)=\mathbf{1}$, set $\mathrm{out}(A)=\zeta(A\to\varepsilon)$ when $A\to\varepsilon$ is a production, and set $\mathrm{out}(A)=\mathbf{0}$ otherwise. A sentential form of a right-linear grammar contains at most one nonterminal, so a derivation of $w=x_1\cdots x_n$ is a sequence $S=A_0\Rightarrow x_1A_1\Rightarrow\cdots\Rightarrow x_1\cdots x_nA_n\Rightarrow w$ closed by $A_n\to\varepsilon$, or a sequence $S=A_0\Rightarrow\cdots\Rightarrow x_1\cdots x_{n-1}A_{n-1}\Rightarrow w$ closed by $A_{n-1}\to x_n$. The first corresponds to the path $A_0\xrightarrow{x_1}A_1\cdots\xrightarrow{x_n}A_n$ with final weight $\zeta(A_n\to\varepsilon)$, and the second to the path through $A_0,\dots,A_{n-1}$ that ends with $A_{n-1}\xrightarrow{x_n}f$ and final weight $\mathbf{1}$. This is a bijection between the derivations of $w$ and the paths labeled by $w$ that start in $S$ and end in $f$ or in a nonterminal with an $\varepsilon$-production, and it preserves the multiset of weights applied. Matched derivations and paths therefore have equal Z-products by~\eqref{eq:derwt} and~\eqref{eq:zbeh}, every other path contributes $\mathbf{0}$, and $\Zlang_G(w)=\beh{\mathcal{A}}(w)$ for every $w$.

Conversely, let $\mathcal{A}$ be a Z-automaton. We first reduce to a single initial state of initial weight $\mathbf{1}$. Add a new state $s$ with $\mathrm{in}(s)=\mathbf{1}$ and $\mathrm{out}(s)=\bigoplus_{q}\mathrm{in}(q)\otimes\mathrm{out}(q)$, add for every letter $x$ and state $q'$ a transition $s\xrightarrow{x}q'$ of weight $\bigoplus_{q}\mathrm{in}(q)\otimes\zeta(q,x,q')$, the sum taken over the states $q$ with $q'\in\Delta(q,x)$, and set the initial weights of the old states to $\mathbf{0}$. By distributivity (Proposition~\ref{prop:semiring}), the paths that leave $s$ carry the Z-sum of the weights of the paths they replace, so the behavior is unchanged. This reduction preserves the behavior but not, in general, a one-to-one correspondence of runs, since one path leaving $s$ stands for the Z-sum of several original paths; the correspondence with derivations established next is one to one for the reduced automaton. Now take the states as nonterminals with start symbol $s$, read each transition $q\xrightarrow{x}q'$ as the production $q\to xq'$ with the same weight, and add the production $q\to\varepsilon$ of weight $\mathrm{out}(q)$ for each state $q$ with $\mathrm{out}(q)\neq\mathbf{0}$. The correspondence of the first part, read backwards, matches the accepting paths labeled by $w$ with the derivations of $w$ and preserves weights, so the resulting right-linear Z-grammar generates $\beh{\mathcal{A}}$.
\end{proof}

\begin{example}[Evidence-derived reliabilities]\label{ex:evid}
One way to give the reliability component content is to estimate it from evidence rather than elicit it. Suppose the grade of a transition is estimated from a corpus, and its reliability is the evidence score $k=n/(n+m)$ of Section~\ref{ssec:znum}, with $m$ a characteristic count. Two Z-automata that share the grade $0.9$ on a decisive transition but estimate it from $n=4$ and from $n=200$ occurrences, against $m=20$, carry the reliabilities $k=4/24\approx0.17$ and $k=200/220\approx0.91$ on that transition, hence, when the remaining factors of the path carry weight $\mathbf{1}$, the credible grades $0.9\cdot0.17\approx0.15$ and $0.9\cdot0.91\approx0.82$ on the acceptance that passes through it. A word whose acceptance uses the transition then receives the same raw grade under both automata but credible grades that differ by roughly fivefold, and the level language $L_{\theta}$ with threshold $\theta=0.5$ admits the word under the well-evidenced automaton and rejects it under the poorly evidenced one. A scalar weight $c=a\,b$ on each transition already reproduces this threshold distinction. What the pair adds is that the two components remain available separately, so that the recognizer can report the reliability of the accepted analysis and filter on it (Section~\ref{ssec:adds}); the evidence-aware representations of our applied work were built to carry the same distinction~\cite{turaev2026}.
\end{example}

\subsection{Discussion and outlook}

Two points orient the development that follows. First, the constructive theory of this section rests on Proposition~\ref{prop:semiring}. Because the Z-weight structure is a commutative semiring, Z-grammars and Z-automata are weighted devices of a standard kind: they inherit the Kleene--Schützenberger characterization and the attendant closure constructions, and the equivalence of Proposition~\ref{prop:equiv} is a weight-preserving instance of the classical one. The credible-grade homomorphism $\cred$ exhibits each Z-device as sitting above an ordinary Viterbi device on the discounted grade, so that the classical constructions transfer along $\cred$ while the two components are carried alongside; decision questions, which the semiring property does not settle, are treated in Section~5. Second, the structure adopted here, selection by the credible grade on the double quotient $\Zzz$, is one principled choice among several. The best-grade variant of Remark~\ref{rem:bestgrade}, the separable and the counting structures, and the native Z-arithmetic are the alternatives. Section~4 examines on which carriers these alternatives satisfy the semiring laws and on which they are only strong bimonoids.

\FloatBarrier
\section{The design space of Z-weight schemes}

The devices of the previous section were defined over one particular way of combining the two components of a Z-weight, the reliability-aware structure $\Zset$ of Definition~\ref{def:zws}, and Remark~\ref{rem:bestgrade} noted that other choices are available and that under some of them the semiring axioms come into question. We now take up that question, since the algebraic class of the weight structure, and not the definition of the device, decides which of the theorems of formal language theory the Z-grammars and Z-automata inherit. We organize the choices along three axes and examine eight representative structures obtained by fixing parameters in these families; they do not exhaust the possible schemes. Seven of them are classified as commutative semirings or as strong bimonoids, the class of the native Z-arithmetic is left unverified, and the consequences for the available theorems are recorded at the end. The positive verdicts are proved. The distributivity laws were in addition checked by exact rational computation over the grid $\{0,\tfrac14,\tfrac12,\tfrac34,1\}^{2}$, and every failure of distributivity that we assert is witnessed by an explicit counterexample drawn from that grid.

\subsection{Three axes and the general family}

A Z-weight is a pair $(a,b)$ in which $a$ is a grade and $b$ a reliability, and three decisions fix how such pairs combine along a derivation and across competing derivations.

\begin{definition}[Z-weight schemes]\label{def:axes}
A \emph{Z-weight scheme} consists of
\begin{enumerate}[label=(\roman*),leftmargin=*,itemsep=0.2ex,topsep=0.3ex]
\item a \emph{grade law} $(S_a,T_a)$, a t-conorm $S_a$ and a t-norm $T_a$ on $[0,1]$;
\item a \emph{path-reliability law} $T_b$, a t-norm on $[0,1]$;
\item an \emph{aggregation scheme} $\sigma$, one of the following:
\begin{enumerate}[label=(\alph*),leftmargin=*,itemsep=0.1ex,topsep=0.1ex]
\item \emph{separable}: grades are aggregated by $S_a$ and reliabilities by a t-conorm $S_b$, independently;
\item \emph{select-by-grade}: the pair of largest grade is reported, ties broken by reliability;
\item \emph{reliability-aware}: the pair of largest credible grade $a\,b$ is reported, ties broken by grade and then by reliability;
\item \emph{native}: Z-numbers are combined by the arithmetic of Section~\ref{ssec:zarith}.
\end{enumerate}
\end{enumerate}
Except under the native scheme, the product is componentwise,
\[
(a_1,b_1)\otimesz(a_2,b_2)=\bigl(T_a(a_1,a_2),\,T_b(b_1,b_2)\bigr).
\]
\end{definition}

The t-conorm $S_a$ aggregates the grades of competing derivations, while $T_a$ and $T_b$ compose grades and reliabilities along a derivation. Scheme~(b) is the best-grade variant of Remark~\ref{rem:bestgrade}, and scheme~(c) is the selection of Definition~\ref{def:zws}, whose last tie-break is vacuous on $\Zzz$.

The three choices do not act independently. Under the separable scheme the reported grade is $S_a$ of the competing grades, and select-by-grade, which presupposes $S_a=\vee$, reports their maximum. Under the reliability-aware scheme the reported grade is that of the most credible derivation, which need not be the largest grade (Example~\ref{ex:ambig}), so this scheme overrides $S_a$ and the grade law contributes only $T_a$.

\begin{definition}[Degenerate-weight quotients]\label{def:quot}
A Z-weight is \emph{degenerate} when the aggregation scheme places it at the bottom of the selection order, that is, when its grade is zero under select-by-grade, and when its grade or its reliability is zero under the reliability-aware scheme. We write
\begin{enumerate}[label=(\roman*),leftmargin=*,itemsep=0.2ex,topsep=0.3ex]
\item $\Zz$ for the quotient of $[0,1]^{2}$ that identifies every weight of grade zero with $\mathbf{0}=(0,0)$, and
\item $\Zzz$ for the quotient of $[0,1]^{2}$ that identifies every weight of grade zero \emph{or} reliability zero with $\mathbf{0}$.
\end{enumerate}
\end{definition}

The quotients express the reading that a derivation which cannot fire, or a grade whose reliability has been discounted to nothing, carries no information and is indistinguishable from the absence of a derivation. Such an identification defines a structure on the quotient only if it is a congruence, that is, if equivalent operands give equivalent sums and equivalent products. Closure of the identified set under multiplication is necessary for this but not sufficient.

\begin{proposition}[Which quotients are congruences]\label{prop:quotcong}
For every grade law and path-reliability law of Definition~\ref{def:axes}:
\begin{enumerate}[label=(\roman*),leftmargin=*,itemsep=0.2ex,topsep=0.3ex]
\item the identification defining $\Zz$ is a congruence under select-by-grade and under the reliability-aware scheme;
\item the identification defining $\Zzz$ is a congruence under the reliability-aware scheme;
\item the identification defining $\Zzz$ is not a congruence under select-by-grade, so this scheme induces no sum on $\Zzz$.
\end{enumerate}
\end{proposition}

\begin{proof}
A t-norm sends $0$ and any argument to $0$, so both identified sets are closed under multiplication by any weight, and the product is compatible with both identifications. For the sum, under select-by-grade a grade-zero weight loses to every weight of positive grade, and two grade-zero weights are both identified with $\mathbf{0}$, which gives the first half of~(i). Under the reliability-aware scheme a weight of credible grade zero loses to every weight of positive credible grade, and two weights of credible grade zero are compared by grade, so that a grade-zero weight loses to every weight of positive grade and two grade-zero weights are both identified with $\mathbf{0}$, which gives the second half of~(i) and gives~(ii). For~(iii), let $s=(1,0)$ and $t=(\tfrac12,\tfrac12)$. Select-by-grade gives $s\oplus t=s$, which is identified with $\mathbf{0}$, whereas $\mathbf{0}\oplus t=t\neq\mathbf{0}$, so equivalent operands produce inequivalent sums, whatever the grade product.
\end{proof}

Accordingly we consider select-by-grade only on $[0,1]^{2}$ and on $\Zz$, and the reliability-aware scheme on $[0,1]^{2}$, $\Zz$, and $\Zzz$.

\subsection{Selection structures and the role of the quotient}

The best-grade variant of Remark~\ref{rem:bestgrade} is a semiring, but only after the degenerate weights are quotiented; on the full carrier it is not one, and establishing exactly where the semiring property holds and where it fails is the first substantive result of the analysis.

\begin{theorem}[Select-by-grade]\label{thm:sel}
With the Viterbi grade law $(\vee,\cdot)$, the path-reliability law $T_b\in\{\cdot,\wedge\}$, and the select-by-grade scheme, the resulting structure is a commutative semiring on the quotient $\Zz$, and on the full carrier $[0,1]^{2}$ it is only a strong bimonoid. The obstruction on the full carrier is the family of grade-zero weights of nonzero reliability, witnessed by
\[
(0,\tfrac14)\otimesz\bigl[(0,\tfrac14)\oplusz(\tfrac14,0)\bigr]=(0,0)
\neq\bigl(0,T_b(\tfrac14,\tfrac14)\bigr)=\bigl[(0,\tfrac14)\otimesz(0,\tfrac14)\bigr]\oplusz\bigl[(0,\tfrac14)\otimesz(\tfrac14,0)\bigr],
\]
where the right-hand side is $(0,\tfrac1{16})$ for $T_b=\cdot$ and $(0,\tfrac14)$ for $T_b=\wedge$.
\end{theorem}

\begin{proof}
By Proposition~\ref{prop:quotcong} the identification defining $\Zz$ is a congruence for both operations, so the structure on $\Zz$ is well defined. On $\Zz$ every element of grade zero is the annihilator $\mathbf{0}$, so the only competing derivations that the sum must weigh are those of strictly positive grade, and there the grade product $a\mapsto a_1a$ is strictly monotone for $a_1>0$, hence order-preserving and injective, so it neither disturbs the maximizer that select-by-grade reports nor creates a tie that was not already present. The grade coordinate of both sides of the distributive law then follows the Viterbi identity $a_1(a_2\vee a_3)=a_1a_2\vee a_1a_3$, and the reliability coordinate on both sides is the $T_b$-product of $b_1$ with the reliability of the common maximizer, so the two sides coincide; commutativity of $\otimesz$ gives distributivity on both sides, and the monoid and annihilation axioms are immediate. On the full carrier the sum is still selection by a total order, the grade refined by the reliability, with least element $\mathbf{0}$, and the product is still componentwise, so the strong-bimonoid axioms hold. Distributivity fails for the displayed triple. On the left, the bracket selects $(\tfrac14,0)$ by its positive grade, and the product is $(0,\tfrac14)\otimesz(\tfrac14,0)=(0,0)$. On the right, the two products are $(0,T_b(\tfrac14,\tfrac14))$ and $(0,0)$; both have grade zero, and the sum returns the first, which is $(0,\tfrac1{16})$ for $T_b=\cdot$ and $(0,\tfrac14)$ for $T_b=\wedge$.
\end{proof}

\begin{theorem}[Reliability-aware selection]\label{thm:seld}
With the Viterbi grade law, the product path-reliability law, and the reliability-aware scheme that selects by the discounted grade $a\,b$, the resulting structure is the commutative semiring $\Zset$ on the double quotient $\Zzz$ (Proposition~\ref{prop:semiring}). On the quotient $\Zz$, which identifies only the grade-zero weights with $\mathbf{0}$, it is a strong bimonoid but not a semiring, the obstruction being the weights of positive grade and reliability zero, witnessed by
\[
(\tfrac14,0)\otimesz\bigl[(\tfrac14,\tfrac14)\oplusz(\tfrac12,0)\bigr]=(\tfrac1{16},0)\neq(\tfrac18,0)=\bigl[(\tfrac14,0)\otimesz(\tfrac14,\tfrac14)\bigr]\oplusz\bigl[(\tfrac14,0)\otimesz(\tfrac12,0)\bigr].
\]
\end{theorem}

\begin{proof}
The semiring property on $\Zzz$ is Proposition~\ref{prop:semiring}. On $\Zz$ the sum selects by a total order, the credible grade refined by the grade and then by the reliability, in which $\mathbf{0}$ is least; the identification defining $\Zz$ is a congruence for the sum and for the product (Proposition~\ref{prop:quotcong}), so both operations are well defined on $\Zz$, and $\mathbf{0}$ annihilates. These are the strong-bimonoid axioms. In the displayed triple, the bracket on the left selects $(\tfrac14,\tfrac14)$, of credible grade $\tfrac1{16}$, over $(\tfrac12,0)$, of credible grade $0$, and the product is $(\tfrac1{16},0)$. On the right both products, $(\tfrac1{16},0)$ and $(\tfrac18,0)$, have credible grade $0$, and the tie is broken by the grade in favor of $(\tfrac18,0)$. Multiplication by a weight of reliability zero thus erases the difference in credible grade that decided the selection on the left, while it keeps the difference in grade that decides it on the right.
\end{proof}

The two selection results isolate the sense in which the construction depends on the arithmetic of the components. The proof of Theorem~\ref{thm:sel} uses only that multiplication by a positive grade is strictly increasing and that $T_b$ is monotone, so its conclusion holds for every strict t-norm in place of the product grade law and for every t-norm $T_b$. The reliability-aware scheme needs more, namely that the credible grade of a product is the product of the credible grades, which holds when both coordinates are multiplied. With $T_b=\wedge$ distributivity fails even on $\Zzz$: for $x=(\tfrac14,\tfrac14)$, $y=(\tfrac14,\tfrac34)$, and $z=(\tfrac12,\tfrac14)$, the left-hand side $x\otimesz(y\oplusz z)$ is $(\tfrac1{16},\tfrac14)$, while the right-hand side is $(\tfrac18,\tfrac14)$. The following proposition shows that the strict monotonicity of the grade law cannot be dropped, so that neither coupled scheme can be built on the minimum on the carriers examined.

\begin{proposition}[The minimum grade law is incompatible with coupled reliability]\label{prop:min}
With the grade law $(\vee,\wedge)$, a path-reliability law $T_b\in\{\cdot,\wedge\}$, and either coupled scheme, select-by-grade or reliability-aware, the resulting structure is a strong bimonoid but not a semiring on the full carrier, on $\Zz$ under either scheme, and on $\Zzz$ under the reliability-aware scheme, which are the carriers on which these schemes are defined (Proposition~\ref{prop:quotcong}). Distributivity fails for $x=(\tfrac14,\tfrac12)$, $y=(\tfrac14,\tfrac12)$, and $z=(\tfrac12,\tfrac14)$:
\[
x\otimesz(y\oplusz z)=\bigl(\tfrac14,\,T_b(\tfrac12,\tfrac14)\bigr)\neq\bigl(\tfrac14,\,T_b(\tfrac12,\tfrac12)\bigr)=(x\otimesz y)\oplusz(x\otimesz z),
\]
that is, $(\tfrac14,\tfrac18)\neq(\tfrac14,\tfrac14)$ for $T_b=\cdot$ and $(\tfrac14,\tfrac14)\neq(\tfrac14,\tfrac12)$ for $T_b=\wedge$. The witness has strictly positive grades and reliabilities throughout, so neither of the degeneracy identifications examined here removes it; this does not exclude that a coarser identification, not considered here, yields a semiring.
\end{proposition}

\begin{proof}
Both coupled sums select by a total order in which $\mathbf{0}$ is least, and the product is componentwise. The minimum sends a weight of grade zero, or of reliability zero, to a weight of grade zero, or of reliability zero, so the argument of Proposition~\ref{prop:quotcong} applies, and the identifications are congruences on the carriers listed. On each of these carriers the two monoid laws and annihilation therefore hold, and the structure is a strong bimonoid. In the bracket on the left both schemes select $z$: select-by-grade by its larger grade, and the reliability-aware scheme because the credible grades of $y$ and $z$ tie at $\tfrac18$ and the tie is broken toward the larger grade. Multiplying by $x$ gives $(\tfrac14,T_b(\tfrac12,\tfrac14))$. On the right the minimum sends both grades to $\tfrac14$, so the two products tie on the grade, and both schemes return $x\otimesz y=(\tfrac14,T_b(\tfrac12,\tfrac12))$, whose reliability, and hence credible grade, is the larger. The minimum is not strictly monotone, so it can merge two distinct grades into a tie, and the two sides of the distributive law then resolve that tie from different reliabilities, before and after the maximization respectively.
\end{proof}

\subsection{Separable structures}

When the reliability is aggregated on its own, the structure decomposes and the semiring property is recovered on the full carrier, at a cost in meaning.

\begin{proposition}[Separable structures are product semirings]\label{prop:sep}
If the grade law $(S_a,T_a)$ and the reliability law $(S_b,T_b)$ are each a commutative semiring on $[0,1]$ and the aggregation is separable, then the resulting structure is their product, hence a commutative semiring on the full carrier $[0,1]^{2}$ with no quotient. In particular $(\vee,\cdot)\times(\vee,\cdot)$ and $(\vee,\cdot)\times(\vee,\wedge)$ are commutative semirings.
\end{proposition}

\begin{proof}
A separable structure is by definition the direct product of the grade structure and the reliability structure, and a finite product of semirings is a semiring, coordinatewise, with units $(\mathbf{1}_{a},\mathbf{1}_{b})$ and $(\mathbf{0}_{a},\mathbf{0}_{b})$; the stated instances are products of a Viterbi semiring with a Viterbi or a fuzzy semiring.
\end{proof}

The price of this clean outcome is that the reliability reported for a word is the best reliability over all of its derivations, aggregated independently of which derivation carries the best grade, so that the structure can return a high reliability that belongs to a weakly graded parse, a decoupling that the selection structures were designed to avoid.

\subsection{Counting structures}

A grade law that sums within the unit interval, rather than selecting, abandons the semiring property for a different reason.

\begin{proposition}[Sum-based grade laws are strong bimonoids]\label{prop:count}
Let the grade law be sum-based, either the probabilistic sum $a+b-ab$ with the product or the bounded sum $\min(a+b,1)$ with the {\L}ukasiewicz t-norm $\max(a+b-1,0)$, and let the reliabilities be aggregated separably by a law $(S_b,T_b)$ that is a commutative semiring on $[0,1]$. Then the resulting structure on $[0,1]^{2}$ is a commutative strong bimonoid and not a semiring.
\end{proposition}

\begin{proof}
Each sum-based grade law is a commutative strong bimonoid on $[0,1]$: its sum is associative and commutative with neutral element $0$, its t-norm is associative and commutative with unit $1$, and $0$ annihilates. The structure is the direct product of this grade bimonoid with the reliability semiring, so it is a commutative strong bimonoid, coordinatewise. Distributivity fails on the grade coordinate. For the probabilistic sum, $a(b+c-bc)-(ab+ac-a^{2}bc)=abc(a-1)$, which is nonzero exactly when $a\notin\{0,1\}$ and $bc\neq0$; for $a=b=c=\tfrac12$ the two sides are $\tfrac38$ and $\tfrac7{16}$. For the bounded sum, $a=b=c=\tfrac12$ gives $\max(\tfrac12+1-1,0)=\tfrac12$ on the left and $\min(0+0,1)=0$ on the right. Giving all three weights the reliability $1$ turns either triple into a witness on the grid. The coupled schemes are not combined with a sum-based grade law, since they select one derivation, whereas a sum-based law accumulates all of them.
\end{proof}

These are the structures under which a word's grade records the combined weight of \emph{all} of its derivations rather than its single best one, the Z-analogue of a probabilistic rather than a Viterbi reading of an ambiguous grammar. The loss of distributivity here is the price of saturation, not of counting as such: the probabilistic and the bounded sums keep the grade inside $[0,1]$, and it is this truncation that breaks the law, whereas the ordinary sum-product structure $([0,\infty),+,\cdot)$ of Table~\ref{tab:semirings} counts and is a semiring. A counting Z-structure over unbounded grades is a possible alternative that we do not examine.

\subsection{The native Z-arithmetic}

The remaining possibility is to combine Z-numbers not by componentwise t-norms but by the native arithmetic of Section~\ref{ssec:zarith}. Aliev et al.~\cite{aliev2015,aliev2016} define the sum and the product of discrete and of continuous Z-numbers, computing the first component by fuzzy arithmetic and the second through the probability distributions compatible with the operands, which are found by optimization and combined by convolution. We do not know of a proof that these operations are associative, and nothing below relies on it. Two features make it a poor fit for the role of $\oplusz$ in the present construction, although neither excludes it from weighted devices in general. The Z-number sum is an \emph{arithmetic} sum: used as the behavior's $\oplusz$ it would add the grades of distinct parses instead of selecting among them, which is a legitimate counting reading, since the probability semiring is not idempotent either, but not the selection reading adopted here. The reliability of a Z-number sum also does not pool the evidence of its summands, since the second component of the sum is formed from the convolution of the underlying distributions and not from the counts on which the summands' reliabilities rest. Because the operations are computed numerically rather than in closed form, we have verified neither the strong-bimonoid axioms for the native pair, namely associativity, a neutral element for the sum, and a zero that annihilates under the product, nor a counterexample to distributivity. Table~\ref{tab:class} accordingly records its class as not verified. Its natural role would be intra-coordinate, in a fuzzy-valued extension of the kind discussed in Remark~\ref{rem:fuzzyvalued}, where the grade and the reliability are themselves fuzzy numbers combined by fuzzy-number arithmetic. Whether any scalar semiring of this section has such a fuzzy-valued counterpart is open (Section~7); selection by a fuzzy credible grade would at least require an order on fuzzy numbers compatible with their product, and Remark~\ref{rem:fuzzyvalued} shows that centroids do not supply one.

\subsection{Classification and consequences}

\begin{figure}[t]
\centering
\includegraphics[width=\textwidth]{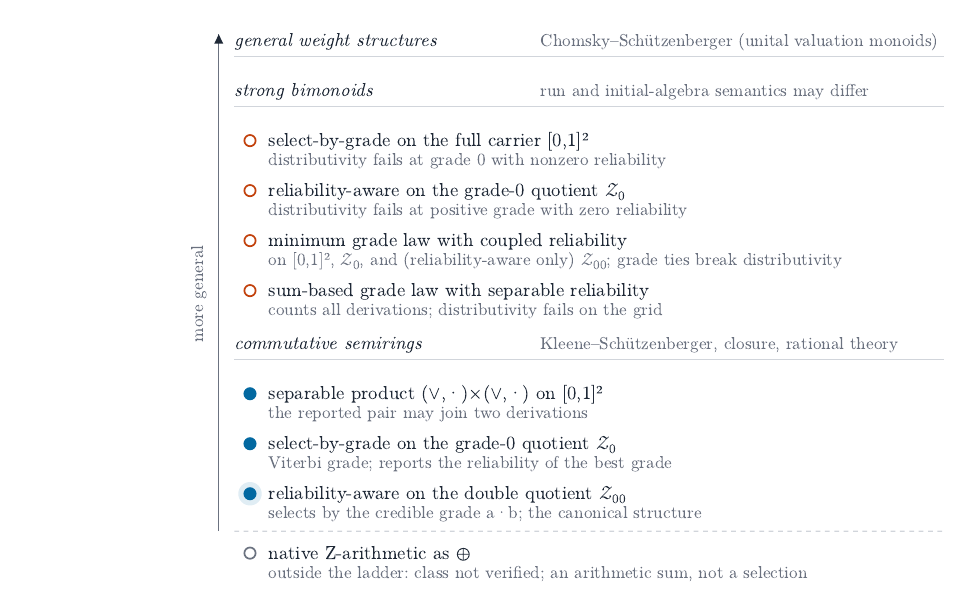}
\caption{The eight representative structures of Table~\ref{tab:class} placed in the hierarchy of weight structures. Three are commutative semirings, on the carriers shown, and retain the full theory. Four are strong bimonoids, which keep the devices and the run semantics but may lose the equality of recognizable and rational series. The native Z-arithmetic, whose class is not verified, is shown outside the ladder. The level labels name the results available at each level under the hypotheses stated in the text.}\label{fig:hier}
\end{figure}

Table~\ref{tab:class} collects the verdicts for the eight representative structures, and Figure~\ref{fig:hier} places them in the hierarchy of weight structures of Section~\ref{ssec:weighted}. Three are commutative semirings, four are strong bimonoids that are not semirings, and the class of the native arithmetic is not verified. The consequence for the theory is read directly from the class. The commutative semirings are the separable product, the select-by-grade structure on $\Zz$, and the reliability-aware structure on $\Zzz$. Each carries the full Kleene--Schützenberger theorem, so that its recognizable and rational Z-languages coincide, together with the closure of the recognizable Z-languages under the rational operations and the equivalence of right-linear Z-grammars with Z-automata of Proposition~\ref{prop:equiv}. A structure that is only a strong bimonoid still defines its Z-automata, and its context-free Z-grammars whenever every word has finitely many derivations of nonzero weight, as it has for grammars without chain productions and $\varepsilon$-productions. Strong bimonoids are unital valuation monoids, and the Chomsky--Sch\"utzenberger theorem of Droste and Vogler~\cite{dv2014} holds for weighted context-free grammars over any unital valuation monoid under this finiteness condition, or over a complete one, so it applies to them. What such a structure can lose is the equality of the recognizable and the rational series, together with the agreement of the run semantics and the initial-algebra semantics of an automaton~\cite{dsv2010}. The correspondence between right-linear Z-grammars and Z-automata survives under the run semantics, because the construction from grammars to automata in the proof of Proposition~\ref{prop:equiv}, and its converse for automata with a single initial state of weight $\mathbf{1}$, match derivations with runs one to one without using distributivity. Such a structure is to be adopted only where this weaker theory is acceptable.

\begin{table}[t]
\centering\footnotesize\setlength{\tabcolsep}{5pt}
\caption{Classification of eight representative structures generated by the Z-weight schemes of Definition~\ref{def:axes}. The eight rows fix parameters in the families of Definition~\ref{def:axes} and do not exhaust them. The semiring and strong-bimonoid verdicts are proved in the text, and every structure classed as a strong bimonoid fails distributivity by a witness on the grid $\{0,\tfrac14,\tfrac12,\tfrac34,1\}^{2}$ given there. Select-by-grade induces no operation on $\Zzz$ (Proposition~\ref{prop:quotcong}), so under the minimum grade law the carrier $\Zzz$ refers to the reliability-aware scheme only. The class of the native arithmetic is not verified.}\label{tab:class}
\begin{tabular}{@{}>{\raggedright\arraybackslash}p{3.5cm}>{\raggedright\arraybackslash}p{1.5cm}>{\raggedright\arraybackslash}p{2.9cm}>{\raggedright\arraybackslash}p{5.4cm}@{}}
\toprule
Structure & Carrier & Class & Theorems kept / semantics\\
\midrule
Separable $(\vee,\cdot)\times(\vee,\cdot)$ & $[0,1]^{2}$ & commutative semiring & full Kleene--Schützenberger; reliability decoupled from the winning grade\\
Select-by-grade, $T_b\in\{\cdot,\wedge\}$ & $\Zz$ & commutative semiring & full Kleene--Schützenberger; selects on grade, reports its reliability\\
Reliability-aware, key $a\,b$ & $\Zzz$ & commutative semiring & full Kleene--Schützenberger; reads grade and reliability together\\
\addlinespace
Select-by-grade, $T_b\in\{\cdot,\wedge\}$ & $[0,1]^{2}$ & strong bimonoid & devices, run semantics, Chomsky--Sch\"utzenberger under finitely many derivations; recognizable and rational series may differ\\
Reliability-aware, key $a\,b$ & $\Zz$ & strong bimonoid & as above; reliability-zero weights break distributivity\\
Minimum grade, coupled reliability & $[0,1]^{2}$, $\Zz$, $\Zzz$ & strong bimonoid & as above; grade ties break distributivity\\
Sum-based grade (count) & $[0,1]^{2}$ & strong bimonoid & as above; accumulates rather than selects\\
\addlinespace
Native Aliev arithmetic & Z-numbers & not verified & not used as $\oplusz$; arithmetic sum, reliability not pooled\\
\bottomrule
\end{tabular}
\end{table}

The classification settles which structure the construction should use. The reliability-aware semiring $\Zset$ on the double quotient $\Zzz$ of Proposition~\ref{prop:semiring} is the main object. It is a genuine commutative semiring, so the full apparatus of weighted automata applies, and its selection by the discounted grade $a\,b$ reads the grade and the reliability together, which the plain select-by-grade rule, by choosing on grade alone, does not honor, as Example~\ref{ex:ambig} shows. The select-by-grade semiring on $\Zz$ of Theorem~\ref{thm:sel} is the simpler baseline, and the separable semiring of Proposition~\ref{prop:sep} is the right choice only where the reliability of a word is to be reported independently of its best parse. The strong bimonoids, among them the counting structures, are to be used only when the weaker theory they support is sufficient, and the native arithmetic is not a candidate for the selection-based $\oplusz$ adopted here. These are the structures on which Section~3 is built, the reliability-aware semiring on $\Zzz$ as the canonical one and the plain select-by-grade structure on $\Zz$ as the baseline, so that the constructive results there rest throughout on a commutative semiring.

\FloatBarrier
\section{Rational theory of Z-languages}

The structure on which the Z-grammars and Z-automata of Section~3 were built is, by Proposition~\ref{prop:semiring}, a commutative semiring. The mature theory of weighted automata over a semiring therefore applies to them without modification, and the purpose of this section is to draw out what that theory yields and, with equal care, where it stops. We record the Kleene--Schützenberger characterization that identifies the recognizable and the rational Z-languages, and the transfer of results along the credible-grade homomorphism of Section~3, which places the Z-theory above the max-product theory of fuzzy automata as a conservative extension. We then establish the closure of the recognizable Z-languages under the rational and the pointwise operations and locate the boundary of decidability. At that boundary the Z-theory inherits the undecidability of equivalence that max-product and tropical automata owe to maximization combined with a product under which finitely many weights generate infinitely many values. Several of these results are standard for commutative, zero-sum-free semirings, and we label them as such; the content specific to Z-weights lies in the reliability filter, in determinization, and in the phenomena of Section~\ref{ssec:adds}. Throughout, $\Zset$ denotes the Z-weight structure of Definition~\ref{def:zws}, the credible-grade semiring on the double quotient $\Zzz$, with credible grade $\cred(a,b)=a\,b$, and $\Vit=([0,1],\vee,\cdot)$ denotes the Viterbi semiring. All decidability and effectivity statements assume rational weights and rational thresholds, on which products and comparisons are exact; the weights estimated from counts in Section~6 are rational when the characteristic count is.

\subsection{Recognizable and rational Z-languages}

A \emph{Z-series}, or \emph{Z-language}, is a map $r\colon\Sigma^{*}\to\Zset$, the set of which forms the semiring $\series{\Zset}$ under pointwise sum and the Cauchy product $(r\cdot s)_{w}=\bigoplus_{uv=w}r_{u}\otimes s_{v}$. A Z-series is \emph{recognizable} when it is the behavior of a Z-automaton in the sense of Definition~\ref{def:zauto}, initial and final Z-weights included, and \emph{rational} when it lies in the least class containing the polynomials, the Z-series of finite support, and closed under the sum, the Cauchy product, and the \emph{star} $r^{*}=\bigoplus_{n\ge0}r^{n}$ taken on \emph{proper} series, those with $r_{\varepsilon}=\mathbf{0}$. The star of a proper series is well defined over any semiring, and in particular over $\Zset$, because the coefficient $(r^{*})_{w}=\bigoplus_{n=0}^{|w|}(r^{n})_{w}$ is a finite Z-sum, a proper series contributing to a word only through derivations of length at most that word.

\begin{theorem}[Kleene--Schützenberger for Z-languages]\label{thm:ks}
A Z-series is recognizable if and only if it is rational. In symbols, $\rec(\Zset)=\rat(\Zset)$.
\end{theorem}

\begin{proof}
The theorem is the specialization to $\Zset$ of a standard result: the equivalence of the recognizable and the rational series holds over an arbitrary semiring~\cite{berstel1988,dk2009,sakarovitch2009,schutz1961}, the two inclusions being the behavior of a rational expression built by the standard automaton constructions for sum, product, and proper star, and the conversion of a Z-automaton into a rational expression by the state-elimination method, whose cycles are handled by the proper star just described. These constructions hold over an arbitrary semiring, so the hypotheses are met for $\Zset$ and both directions apply verbatim; commutativity, which $\Zset$ also enjoys (Proposition~\ref{prop:semiring}), is not required for this theorem but is used for the Hadamard product below.
\end{proof}

\begin{remark}[Cyclic weights and the best-path reading]\label{rem:star}
For a single weight $z=(a,b)$ the constant series $z\varepsilon$ is not proper unless $z=\mathbf{0}$, so its star lies outside the rational operations above, and the infinite Z-sum $\bigoplus_{n\ge0}z^{n}$ has to be read directly. The powers $z^{n}=(a^{n},b^{n})$ have credible grades $(ab)^{n}\le1$, and the term $z^{0}=\mathbf{1}=(1,1)$ is the greatest element of $\Zset$ in the order of~\eqref{eq:zoplus}, being the only Z-weight of credible grade $1$. The family therefore has a greatest element, the sum is attained at $n=0$, and $\bigoplus_{n\ge0}z^{n}=\mathbf{1}$. The star of a proper series needs no such reading, since each of its coefficients is the finite Z-sum described above. In the same way, since every weight is at most $\mathbf{1}$, a path that traverses a cycle has a weight no greater than that of the same path with the cycle removed. When the cycle is labeled by a nonempty word the shortened path carries a different word, so this compares path weights irrespective of their labels, as in the state-elimination construction, and not the paths for a single word. For a fixed word the behavior is carried by its most credible accepting path, the Z-analogue of the shortest-path reading of a tropical automaton and the best-path reading of a Viterbi automaton. This is the structural reason that the behaviors of Section~3 were computed by a single path.
\end{remark}

\subsection{Transfer along the credible grade}

The credible-grade map $\cred$ of Section~3 relates the Z-theory to the ordinary max-product theory of fuzzy automata, and the relation is a retraction, so that the Z-theory extends the fuzzy theory without disturbing it.

\begin{proposition}[Conservative extension]\label{prop:transfer}
The credible grade $\cred\colon\Zset\to\Vit$ and the reliability-one embedding $\iota\colon\Vit\to\Zset$, $\iota(a)=(a,1)$, are semiring homomorphisms with $\cred\circ\iota=\mathrm{id}_{\Vit}$. Applied coefficientwise they induce semiring homomorphisms $\hat{\cred}\colon\series{\Zset}\to\series{\Vit}$ and $\hat{\iota}\colon\series{\Vit}\to\series{\Zset}$ that commute with the rational operations and send recognizable series to recognizable series; consequently the credible grade of a recognizable Z-language is a recognizable max-product fuzzy language, and every recognizable max-product fuzzy language is the credible grade of a recognizable Z-language in which every weight is fully reliable.
\end{proposition}

\begin{proof}
That $\cred$ is a homomorphism is Proposition~\ref{prop:semiring}, and $\iota$ is a homomorphism because $(a,1)\otimes(a',1)=(aa',1)$ and the credible grade of $(a,1)$ is $a$, so the selection in $\oplus$ agrees with the maximum on $\Vit$; the composite $\cred\circ\iota$ is the identity since $\cred(a,1)=a$. A semiring homomorphism applied to the weights of a weighted automaton yields a weighted automaton whose behavior is the image series, so $\hat{\cred}$ and $\hat{\iota}$ preserve recognizability, and they commute with the rational operations because these are defined by the semiring operations that the homomorphisms preserve. The two stated consequences are the images under $\hat{\cred}$ and $\hat{\iota}$ of a recognizable series, together with $\hat{\cred}\circ\hat{\iota}=\mathrm{id}$.
\end{proof}

\begin{figure}[t]
\centering
\includegraphics[width=0.7\textwidth]{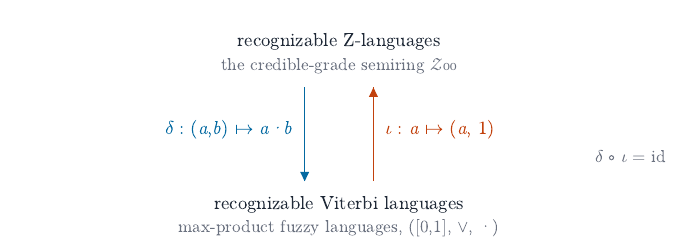}
\caption{The credible grade $\cred$ and the reliability-one embedding $\iota$ form a retraction of the Z-theory onto the max-product fuzzy theory. Both preserve recognizability and the rational operations, and $\cred\circ\iota$ is the identity, so the Z-theory is a conservative extension.}\label{fig:transfer}
\end{figure}

Figure~\ref{fig:transfer} records the retraction. Its force is that nothing of the max-product fuzzy theory is lost in passing to Z-automata: a fuzzy automaton is the reliability-one Z-automaton whose reliabilities have been set aside, and its Z-refinement differs only in carrying the second component that the credible grade would otherwise discard.

\begin{example}[Projecting the running automaton]\label{ex:proj}
The Z-automaton of Example~\ref{ex:zauto} has transition weights $(0.9,0.9)$, $(0.5,0.6)$, and $(0.7,0.8)$, whose credible grades are $0.81$, $0.30$, and $0.56$. Its image under $\hat{\cred}$ is the Viterbi automaton on the same skeleton with these three grades, whose behavior on $aab$ is $0.81\cdot0.30\cdot0.56=0.13608\approx0.136$, the credible grade of the Z-behavior $(0.315,0.432)$ computed in Section~3. The Z-automaton refines this scalar by recording that the accepting path has grade $0.315$ and reliability $0.432$, the information that the credible grade alone cannot separate.
\end{example}

\subsection{Closure properties}

\begin{proposition}[Closure]\label{prop:closure}
The recognizable Z-languages are closed under the sum $\oplus$, the Cauchy product, the proper star, the scalar products $z\otimes r$ and $r\otimes z$ for $z\in\Zset$, the Hadamard (pointwise) product $(r\odot s)_{w}=r_{w}\otimes s_{w}$, the reversal, and the application of a nonerasing alphabet morphism and of an inverse alphabet morphism.
\end{proposition}

\begin{proof}
These closure properties are standard for weighted automata over a commutative semiring~\cite{dk2009,sakarovitch2009}, and we indicate the constructions only to show where commutativity enters. Closure under the rational operations is Theorem~\ref{thm:ks}, and the scalar products are the Cauchy products with the constant series $z\varepsilon$, which is recognized by a one-state automaton with initial weight $z$, final weight $\mathbf{1}$, and no letter transitions. The Hadamard product is recognized by the product automaton whose initial, transition, and final weights are the $\otimes$-products of the paired weights; this construction preserves behavior precisely because $\Zset$ is commutative, so that the weight of a pair of paths is the $\otimes$-product of their weights irrespective of order, which is where Proposition~\ref{prop:semiring} is used. Reversal is recognized by the reversed automaton with the initial and final weights exchanged, again by commutativity. For an alphabet morphism $h\colon\Sigma^{*}\to\Gamma^{*}$ that is nonerasing, $h(x)\neq\varepsilon$ for every $x\in\Sigma$, the image of $r$ is the series $h(r)$ over $\Gamma$ with $(h(r))_{v}=\bigoplus_{w\in h^{-1}(v)}r_{w}$, a finite Z-sum because $|w|\le|v|$ for every $w\in h^{-1}(v)$. It is recognized by replacing each transition $q\xrightarrow{x}q'$ of weight $z$ by a chain of fresh states spelling $h(x)=y_1\cdots y_k$, whose first transition carries $z$ and whose remaining transitions carry $\mathbf{1}$; the paths of the new automaton labeled $v$ correspond one to one, with equal weights, to the paths of the old one labeled by some $w\in h^{-1}(v)$. For a morphism $g\colon\Gamma^{*}\to\Sigma^{*}$, erasing or not, the inverse image of $r$ is the series $g^{-1}(r)$ over $\Gamma$ with $(g^{-1}(r))_{v}=r_{g(v)}$. Write the automaton as an initial row vector $\mathrm{in}$, transition matrices $\mu(x)\in Z^{Q\times Q}$, and a final column vector $\mathrm{out}$, so that $r_{w}=\mathrm{in}\,\mu(w)\,\mathrm{out}$, with $\mu$ extended multiplicatively to words and $\mu(\varepsilon)$ the identity matrix. The automaton with the same $\mathrm{in}$ and $\mathrm{out}$ and the matrices $\mu'(y)=\mu(g(y))$ for $y\in\Gamma$ then recognizes $g^{-1}(r)$, a letter $y$ with $g(y)=\varepsilon$ receiving the identity matrix; the matrix products are defined over $\Zset$ because every entry is a finite Z-sum~\cite{sakarovitch2009}.
\end{proof}

\subsection{Supports, cut-points, and decidability}

The qualitative content of a Z-language, the set of words it generates at all, is regular, and so are its quantitative cuts, because grades and reliabilities never exceed one; what inherits the hardness of max-product automata is the comparison of two Z-languages as a whole.

\begin{proposition}[The support is regular]\label{prop:support}
The \emph{support} $\operatorname{supp}(r)=\{w:r_{w}\neq\mathbf{0}\}$ of a recognizable Z-language is a regular language, and its emptiness is decidable.
\end{proposition}

\begin{proof}
The semiring $\Zset$ is zero-sum-free, since the $\oplus$ of weights of positive credible grade selects one of them and so is never $\mathbf{0}$, and it has no zero divisors, since $(a_1,b_1)\otimes(a_2,b_2)=\mathbf{0}$ on $\Zzz$ forces a factor of credible grade $0$, that is a factor $\mathbf{0}$. Over a zero-sum-free semiring without zero divisors the support of a recognizable series is the language of the underlying Boolean automaton, obtained by sending every nonzero weight, initial and final weights included, to $1$, hence regular, and its emptiness reduces to reachability of an accepting state, which is decidable; this is standard~\cite{dk2009,sakarovitch2009}.
\end{proof}

\begin{proposition}[Level and reliability-filtered languages are regular]\label{prop:cut}
For every Z-automaton $\mathcal{A}$ with rational weights, deterministic or not, every rational $\theta\in(0,1]$, and every rational $\rho\in(0,1]$, the level language and the reliability-filtered language
\[
L_{\theta}(\mathcal{A})=\{w:\cred(\beh{\mathcal{A}}(w))\ge\theta\},\qquad
L_{\theta,\rho}(\mathcal{A})=\{w:\cred(\beh{\mathcal{A}}(w))\ge\theta\ \text{and}\ \rel(\beh{\mathcal{A}}(w))\ge\rho\}
\]
are regular, and finite automata recognizing them are effectively constructible from $\mathcal{A}$, $\theta$, and $\rho$. In particular, the emptiness and the universality of these languages are decidable.
\end{proposition}

\begin{proof}
Call a \emph{factor} of $\mathcal{A}$ any of its initial, transition, and final weights, so that the weight of an accepting path is the Z-product of its initial weight, its transition weights, and its final weight. Since grades and reliabilities lie in $[0,1]$, a factor whose credible grade equals $1$ is $\mathbf{1}=(1,1)$; we call every factor other than $\mathbf{0}$ and $\mathbf{1}$ \emph{costly}. If $\mathcal{A}$ has no costly factor, every accepting path of nonzero weight has weight $\mathbf{1}$, so the behavior is $\mathbf{1}$ on the support and $\mathbf{0}$ off it, and since $\theta,\rho>0$ both languages equal the support, which is regular by Proposition~\ref{prop:support}. Otherwise let $\lambda$ be the largest credible grade of a costly factor, so that $0<\lambda<1$, and let $\ell$ be the largest integer $k\ge0$ with $\lambda^{k}\ge\theta$, which equals $\lfloor\log\theta/\log\lambda\rfloor$ and, for rational $\lambda$ and $\theta$, is computed exactly by forming the powers $\lambda,\lambda^{2},\dots$ until one falls below $\theta$. For $\theta=1$ this gives $\ell=0$, so that only paths all of whose factors equal $\mathbf{1}$ are retained, as they must be. Because $\cred$ is multiplicative and no factor has credible grade above $1$, an accepting path of credible grade at least $\theta$ involves at most $\ell$ costly factors, and so does each of its prefixes, whose credible grade is at least as large. Its weight, and the weight of each prefix, therefore lies in the finite set $W_{\theta}$ of Z-products of at most $\ell$ costly factors whose credible grade is at least $\theta$, the empty product $\mathbf{1}$ included. Let $\mathcal{B}$ be the nondeterministic finite automaton on the states $Q\times W_{\theta}$ whose initial states are the pairs $(q,u)$ with $u$ the initial weight of $q$ and $u\in W_{\theta}$, and whose transitions $(q,u)\xrightarrow{x}(q',u\otimes z)$ correspond to the transitions $q\xrightarrow{x}q'$ of weight $z\neq\mathbf{0}$ with $u\otimes z\in W_{\theta}$. On input $w$, $\mathcal{B}$ reaches a set $S(w)$ of states, and its subset automaton is finite. Let $F(w)$ be the set of products $u\otimes t$ with $(q,u)\in S(w)$, $t$ the final weight of $q$, and $\cred(u\otimes t)\ge\theta$; it is exactly the set of weights of credible grade at least $\theta$ of the accepting paths for $w$. A word $w$ lies in $L_{\theta}(\mathcal{A})$ exactly when $F(w)\neq\emptyset$. In that case the behavior $\beh{\mathcal{A}}(w)$, which by~\eqref{eq:zoplus} is the greatest accepting path weight in the order by credible grade, then grade, is the greatest element of $F(w)$, since a path of credible grade below $\theta$ is never selected over one of credible grade at least $\theta$. Membership in either language is therefore decided by $F(w)$, which depends only on $S(w)$, so the subset automaton with the corresponding accepting sets recognizes both languages, and it is computed from finitely many products of factors.
\end{proof}

\begin{remark}[Why the maximization does not obstruct thresholds]\label{rem:cut}
For scalar weights, Proposition~\ref{prop:cut} is folklore for automata with nonnegative costs, the distance automata treated by Kuich~\cite{kuich1997}; the content specific to Z-weights is that the selected pair, and hence the reliability filter, is decided by the finite set $W_{\theta}$ of products. The level languages are the cut-point languages of the Viterbi image $\hat{\cred}(r)$, and it might be expected that they inherit the undecidability results for max-plus automata. They do not, because the Viterbi semiring corresponds, through the map $x\mapsto-\log x$, to the min-plus semiring with \emph{nonnegative} costs rather than to a semiring with costs of both signs, and for nonnegative costs a constant threshold bounds the number of costly steps, which is what the proof of Proposition~\ref{prop:cut} exploits. A finite catalog of weights does generate infinitely many products under $\otimes$, but only finitely many of them reach a positive threshold. For distance automata even the question whether the costs of all accepted words are bounded is decidable~\cite[Theorem~10.16]{kuich1997}. The undecidability result of Krob~\cite{krob1994} is stated over the tropical semiring $(\mathbb{N}\cup\{\infty\},\min,+)$ and transfers to the Viterbi semiring through the injective semiring homomorphism $k\mapsto2^{-k}$, $\infty\mapsto0$. It concerns the comparison of an automaton with another automaton or with a fixed series, not with a constant, and it survives in Proposition~\ref{prop:dec}. The construction is exponential in $\ell$, which grows as $\log\theta/\log\lambda$, so it is effective but not efficient when the costly weights lie close to $\mathbf{1}$. We checked it against exhaustive path enumeration on sixty random nondeterministic Z-automata with three states over a two-letter alphabet, on every word of length at most seven, for both languages.
\end{remark}

\begin{proposition}[The boundary of decidability]\label{prop:dec}
Emptiness of the support of a Z-automaton is decidable. Behavioral equivalence of two Z-automata is undecidable in general, but for Z-automata with rational weights it is decidable in polynomial time when both are unambiguous, and in particular when both are deterministic.
\end{proposition}

\begin{proof}
The first assertion is Proposition~\ref{prop:support}. Undecidability is obtained by reduction from the equivalence of tropical automata over $(\mathbb{N}\cup\{\infty\},\min,+)$, which is undecidable~\cite{krob1994}. The encoding $k\mapsto2^{-k}$ of Remark~\ref{rem:cut}, followed by the embedding $\iota(a)=(a,1)$ of Proposition~\ref{prop:transfer}, is an injective semiring homomorphism into $\Zset$ with rational values. Applying it to the weights of a tropical automaton yields a Z-automaton whose behavior is the image of the original, so an equivalence test for Z-automata would decide the equivalence of their tropical preimages.

For the positive assertion, let both Z-automata be unambiguous, so that each word labels at most one accepting path in each. The behavior on a word is then the pair of the product of the grades and the product of the reliabilities along that path, initial and final weights included, with no selection, and $\mathbf{0}$ when there is no such path. Hence the grade image $\grade\circ\beh{\mathcal{A}}$ is computed by the automaton over $([0,1],\cdot)$ that keeps the grade of every weight, and because each word has at most one accepting path, the maximum over paths may be replaced by the sum, so that the same automaton read over the field $\mathbb{Q}$ of rationals computes the same function. The same holds for the reliability image $\rel\circ\beh{\mathcal{A}}$. Since the nonzero elements of $\Zzz$ have positive components, two unambiguous Z-automata are equivalent exactly when their grade images agree and their reliability images agree, and equivalence of $\mathbb{Q}$-weighted automata is decidable in polynomial time~\cite{schutz1961,tzeng1992}. A deterministic automaton is unambiguous, so the deterministic case is a special case and needs no minimal forms.
\end{proof}

\subsection{Determinization and minimization}

Following Mohri~\cite{mohri1997}, two states $p$ and $p'$ of a Z-automaton are \emph{siblings} when there are words $x$ and $y$ such that both are reached from states of nonzero initial weight by paths labeled $x$ and both carry a cycle labeled $y$, and the automaton has the \emph{twins property} when, for all siblings $p,p'$ and all such $y$, the Z-sum of the weights of the cycles labeled $y$ at $p$ equals that at $p'$.

\begin{proposition}[Determinization and minimal deterministic form]\label{prop:det}
Let $\widehat{\Zset}$ be the extension of $\Zset$ to the carrier $\{\mathbf{0}\}\cup(0,\infty)^{2}$, with the product~\eqref{eq:zotimes} and the order of Definition~\ref{def:zws}.
\begin{enumerate}[label=(\roman*),leftmargin=*,itemsep=0.2ex,topsep=0.3ex]
\item Whenever the weighted subset construction terminates on a Z-automaton, it returns a deterministic automaton over $\widehat{\Zset}$ equivalent to it; the weights of this automaton may lie outside $[0,1]^{2}$, although its behavior takes values in $\Zset$.
\item On a trim unambiguous Z-automaton with the twins property the construction terminates.
\item A trim deterministic automaton over $\widehat{\Zset}$ whose behavior takes values in $\Zset$ has a minimal equivalent deterministic automaton over $\widehat{\Zset}$, unique up to an isomorphism of states and weight pushing.
\end{enumerate}
\end{proposition}

\begin{proof}
A weight of positive credible grade is cancellable under $\otimes$, which is coordinatewise multiplication, but the residual $(a_1,b_1)^{-1}\otimes(a_2,b_2)=(a_2/a_1,\,b_2/b_1)$ may have a coordinate exceeding $1$. The pair $(0.9,0.3)$ against the selected $(0.6,0.5)$ already has residual $(1.5,0.6)$, and a later transition of weight $(0.9,0.9)$ out of the corresponding subset state then receives the weight $(1.35,0.54)$. The construction is therefore carried out in $\widehat{\Zset}$. Through the injective map $(a,b)\mapsto(ab,a)$ of Section~\ref{ssec:zws}, the nonzero elements of $\widehat{\Zset}$ form the multiplicative group $(0,\infty)^{2}$, and the order of~\eqref{eq:zoplus}, by credible grade and then by grade, becomes the lexicographic order on this group, which is total and compatible with the product. Hence $\widehat{\Zset}$ is a commutative selective semiring, zero-sum-free, with every nonzero element invertible, so it is cancellative and divisible, and $\Zset$ is a subsemiring of it.

(i) The weighted subset construction~\cite{mohri1997,mohri2009} builds states that are finite sets of pairs $(q,r)$ of a state and a nonzero residual weight. On a word $u$ it reaches the subset $S_u$ with the output weight $\lambda(u)$, where $\lambda(u)$ is the Z-sum of the weights of all paths labeled $u$ from states of nonzero initial weight, initial weights included, and the residual of $q$ is $r_u(q)=\lambda(u)^{-1}\otimes c_u(q)$, with $c_u(q)$ the Z-sum of the weights of those paths that end in $q$. The invariant $\lambda(u)\otimes r_u(q)=c_u(q)$ holds for $u=\varepsilon$ by definition and is preserved by each step, because the step to $ux$ computes $c_{ux}(q')=\bigoplus_{q}c_u(q)\otimes\zeta(q,x,q')$ by distributivity and divides by the Z-sum of these values, which is possible because nonzero elements of $\widehat{\Zset}$ are invertible. The final weight of $S_u$ is $\bigoplus_{q}r_u(q)\otimes\mathrm{out}(q)$, so the output on $u$ is $\lambda(u)\otimes\bigoplus_q r_u(q)\otimes\mathrm{out}(q)=\bigoplus_q c_u(q)\otimes\mathrm{out}(q)=\beh{\mathcal{A}}(u)$. Hence the construction, whenever it terminates, returns an equivalent deterministic automaton, whose weights need not lie in $[0,1]^{2}$.

(ii) Let $\mathcal{A}$ be trim and unambiguous. For every word $u$ and state $q$ there is at most one path labeled $u$ from a state of nonzero initial weight to $q$, since two such paths, extended by a common path from $q$ to a final state, would give two accepting paths for one word; for the same reason a state carries at most one cycle with a given label. Let $\pi_q$ be the path to $q\in S_u$, and $\pi_{*}$ the one whose weight $\lambda(u)$ selects, so that $r_u(q)=\operatorname{wt}(\pi_{*})^{-1}\otimes\operatorname{wt}(\pi_q)$, initial weights included. If $|u|\ge|Q|^{2}$, the pairs of states visited by $\pi_q$ and $\pi_{*}$ at the $|u|+1$ positions repeat, so there are positions $i<j$ at which both paths visit the same pair $(p,p')$. The segments between them are cycles $c$ at $p$ and $c'$ at $p'$ with a common label $y$, and $p$ and $p'$ are siblings, both being reached by the prefix of $u$ of length $i$. By the twins property and the uniqueness of cycles with a given label, $\operatorname{wt}(c)=\operatorname{wt}(c')$. Removing both cycles yields paths $\sigma_q$ and $\sigma_{*}$ with a common label of length $|u|-|y|$, and commutativity and cancellation give $\operatorname{wt}(\sigma_{*})^{-1}\otimes\operatorname{wt}(\sigma_q)=r_u(q)$. The argument applies again to $\sigma_q$ and $\sigma_{*}$, since it uses only that the two paths start at states of nonzero initial weight and carry a common label, so after finitely many removals $r_u(q)$ is written as $\operatorname{wt}(\sigma)^{-1}\otimes\operatorname{wt}(\sigma')$ for two paths with a common label of length below $|Q|^{2}$. There are finitely many such pairs, hence finitely many residuals and finitely many subset states, and the construction terminates.

(iii) Let $\mathcal{D}$ be a trim deterministic automaton over $\widehat{\Zset}$ whose behavior takes values in $\Zset$. Every cycle of $\mathcal{D}$ has its weight in $(0,1]^{2}$, since otherwise pumping the cycle would push a component of the behavior above $1$, and such a weight is at most $\mathbf{1}$ in the order of Definition~\ref{def:zws}. Removing a cycle from a path to a final state therefore never lowers its weight, so the Z-sum $d(q)$ of the weights of the paths from $q$ to a final state, final weight included, is attained on a path without repeated states, and it is nonzero because $\mathcal{D}$ is trim. \emph{Weight pushing} replaces the initial weight $\alpha$ by $\alpha\otimes d(q_0)$, each transition weight $\zeta(q,x,q')$ by $d(q)^{-1}\otimes\zeta(q,x,q')\otimes d(q')$, and each final weight $\mathrm{out}(q)$ by $d(q)^{-1}\otimes\mathrm{out}(q)$; the products along a path telescope, so the behavior is unchanged, and in the pushed automaton every state has $d(q)=\mathbf{1}$, which is the residual-state normalization. Two states of the pushed automaton have the same future behavior exactly when they are equivalent as states of the unweighted automaton over the pairs of a letter and a weight, and minimizing that automaton, whose alphabet also records the final weights, gives a minimal deterministic automaton by the construction of Mohri~\cite{mohri2009}, which applies because $\widehat{\Zset}$ is divisible and zero-sum-free. Two minimal deterministic automata with the same behavior become isomorphic after pushing, since pushing normalizes the residual of each state and the normalized residuals determine the states, so the minimal form is unique up to an isomorphism of states and weight pushing; the pushed weights again need not lie in $[0,1]^{2}$.
\end{proof}

For unambiguous Z-automata, the twins property on the \emph{Z-weights} thus suffices for determinization. This property implies the twins property of the credible-grade image because $\cred$, extended to $\widehat{\Zset}$, is a homomorphism. For ambiguous Z-automata we do not claim that the twins property suffices. Through $\eta$ and coordinatewise logarithms, $\widehat{\Zset}$ is isomorphic to the max-plus semiring over the lexicographically ordered group $\mathbb{R}^{2}$, and Mohri's termination argument for the tropical semiring may well extend to it, but the extension requires a proof that we do not give here (Section~7). The converse fails: two cycles may carry the same credible grade $a\,b$ while differing in grade and in reliability, so that the Z-weights fail the twins property although their credible grades satisfy it. The next proposition shows that the failure is not an artifact of the construction. An unambiguous Z-automaton may have no deterministic equivalent at all, over $\widehat{\Zset}$ or over any commutative semiring that contains $\Zset$, although its Viterbi image determinizes, so the twins property on the Z-weights is strictly stronger than on their credible grades.

\begin{proposition}[A Z-automaton without a deterministic equivalent]\label{prop:nondet}
Let $\mathcal{N}$ be the Z-automaton over $\{a,c,d\}$ with the states $q_0,p,r,f$, the initial state $q_0$ and the final state $f$, both with weight $\mathbf{1}$, and the transitions
\[
\begin{gathered}
q_0\xrightarrow{a}p:(\tfrac12,1),\quad q_0\xrightarrow{a}r:(1,\tfrac12),\quad p\xrightarrow{a}p:(\tfrac12,1),\\
r\xrightarrow{a}r:(1,\tfrac12),\quad p\xrightarrow{c}f:\mathbf{1},\quad r\xrightarrow{d}f:\mathbf{1}.
\end{gathered}
\]
Then $\mathcal{N}$ is unambiguous, with $\beh{\mathcal{N}}(a^{n}c)=(2^{-n},1)$ and $\beh{\mathcal{N}}(a^{n}d)=(1,2^{-n})$ for $n\ge1$ and $\beh{\mathcal{N}}(w)=\mathbf{0}$ for every other word $w$. Its Viterbi image $\hat{\cred}(\beh{\mathcal{N}})$ is the behavior of a deterministic Viterbi automaton, but no deterministic automaton over a commutative semiring that contains $\Zset$ as a subsemiring, in particular over $\widehat{\Zset}$, with initial and final weights allowed, has the behavior $\beh{\mathcal{N}}$.
\end{proposition}

\begin{proof}
A word accepted by $\mathcal{N}$ leaves $q_0$ on $a$, stays in $p$ or in $r$ while reading $a$, and enters $f$ on $c$ from $p$ or on $d$ from $r$, so the accepted words are $a^{n}c$ and $a^{n}d$ with $n\ge1$, each with exactly one accepting path. The path for $a^{n}c$ takes $n$ transitions of weight $(\tfrac12,1)$ and then $\mathbf{1}$, and the path for $a^{n}d$ takes $n$ transitions of weight $(1,\tfrac12)$ and then $\mathbf{1}$, which gives the stated behavior. Both values have credible grade $2^{-n}$, which is the behavior of the deterministic Viterbi automaton with the transitions $q_0\xrightarrow{a}s$ and $s\xrightarrow{a}s$ of weight $\tfrac12$ and $s\xrightarrow{c}f$ and $s\xrightarrow{d}f$ of weight $1$.

Suppose that a deterministic automaton $\mathcal{D}$ over a commutative semiring $\Sem\supseteq\Zset$, with initial weight $\alpha$ on its initial state and arbitrary final weights, has the behavior $\beh{\mathcal{N}}$. A deterministic automaton assigns to each word the weight of its unique path, preceded by $\alpha$ and followed by the final weight of the last state, and the empty sum $\mathbf{0}$ when that path does not exist. Since $a^{n}c$ has a nonzero value for every $n\ge1$, the path on $a^{n}$ exists; let $s_n$ be the state it reaches and $u_n\in\Sem$ the product of $\alpha$ and its transition weights. As $\mathcal{D}$ has finitely many states, there are $1\le n<n'$ with $s_n=s_{n'}=s$. Let $x_c$ be the product of the weight of the $c$-transition leaving $s$ and the final weight of its target, and define $x_d$ likewise; both transitions exist, since $a^{n}c$ and $a^{n}d$ have nonzero values. Then
\[
u_n\otimes x_c=(2^{-n},1),\quad u_n\otimes x_d=(1,2^{-n}),\quad u_{n'}\otimes x_c=(2^{-n'},1),\quad u_{n'}\otimes x_d=(1,2^{-n'}).
\]
By commutativity and associativity, $(u_n\otimes x_c)\otimes(u_{n'}\otimes x_d)=(u_{n'}\otimes x_c)\otimes(u_n\otimes x_d)$, and since the four factors lie in $\Zset$, which is closed under $\otimes$ in $\Sem$, this reads $(2^{-n},2^{-n'})=(2^{-n'},2^{-n})$ in $\Zset$. Both pairs are nonzero elements of $\Zzz$, so $n=n'$, a contradiction.
\end{proof}

In the terms of Mohri~\cite{mohri1997}, the states $p$ and $r$ of $\mathcal{N}$ are siblings, both reached by $a$ and both carrying a cycle on $a$, whose cycle weights $(\tfrac12,1)$ and $(1,\tfrac12)$ differ although their credible grades coincide. The reliabilities are thus not merely carried along; they enter the determinizability condition in their own right.

\subsection{What the reliability component adds}\label{ssec:adds}

Because the credible grade is a semiring homomorphism, every statement about a Z-automaton that is phrased through the credible grade alone is a statement about its Viterbi image. The level language $L_{\theta}(\mathcal{A})$ is the cut-point language of $\hat{\cred}(\mathcal{A})$, and Proposition~\ref{prop:transfer} shows that the credible-grade part of the theory is the max-product theory read on the discounted grade. A reader may therefore ask what the second component contributes. The answer is concentrated in three places: the reliability filter, its behavior when analyses are added, and determinization. Below, $|w|_{x}$ denotes the number of occurrences of the letter $x$ in $w$.

\begin{proposition}[Reliability filters are not scalar thresholds]\label{prop:notscalar}
Let $\mathcal{A}$ be the one-state Z-automaton over $\{a,b\}$ whose state is initial and accepting, with the loop on $a$ weighted $(\tfrac12,1)$ and the loop on $b$ weighted $(1,\tfrac12)$, and let $\theta=2^{-6}$ and $\rho=2^{-3}$. Then
\[
L_{\theta,\rho}(\mathcal{A})=\{w:|w|\le6,\ |w|_{b}\le3\},
\]
and no assignment of positive real weights to the two transitions, with any initial and final weights, any constant threshold, and any of the relations $<,\le,=,\ne,\ge,>$, makes the scalar max-product automaton on the same one-state skeleton recognize this language.
\end{proposition}

\begin{proof}
The automaton has a single run on each word $w$, of weight $(2^{-|w|_{a}},2^{-|w|_{b}})$, with credible grade $2^{-|w|}$ and reliability $2^{-|w|_{b}}$, which gives the displayed language. A scalar reweighting with transition weights $e^{-x}$ and $e^{-y}$, its initial and final weights absorbed into the threshold, yields the language of the words with $x|w|_{a}+y|w|_{b}\bowtie C$ for real $x$, $y$, and $C$, where $\bowtie$ stands for the relation in question. The set $R=\{(i,j):i+j\le6,\ j\le3\}$ of letter counts of the target is finite and contains the noncollinear points $(0,0)$, $(1,0)$, and $(0,1)$. This excludes $=$, since the solutions in $\mathbb{N}^{2}$ of $xi+yj=C$ lie on a line when $(x,y)\neq(0,0)$ and form either the empty set or all of $\mathbb{N}^{2}$ otherwise. It also excludes $\ne$, since the solutions of $xi+yj\ne C$ form the complement in $\mathbb{N}^{2}$ of such a set, which is either empty or infinite, whereas $R$ is finite and nonempty. The relations $\ge$ and $>$ reduce to $\le$ and $<$ by negating $x$, $y$, and $C$. Suppose that $xi+yj\le C$ holds exactly on $R$. From $(6,0)\in R$ and $(7,0)\notin R$ we obtain $6x\le C<7x$, so $x>0$, $C>0$, and $x>C/7$, and from $(0,3)\in R$ and $(0,4)\notin R$ we obtain $y>C/4$. Then $(3,3)\in R$ requires $3x+3y\le C$, whereas $3x+3y>\tfrac{3}{7}C+\tfrac{3}{4}C=\tfrac{33}{28}C>C$. The strict relation is excluded by the same argument, and an exact linear-programming check over all letter counts up to eleven confirms that no separating half-plane exists in either orientation.
\end{proof}

The language is finite and therefore recognized by some other scalar automaton, so the proposition concerns information rather than generative capacity. For a given Z-automaton, the question which words are admitted with an analysis that is both credible and reliable cannot be delegated to one weight per transition, whatever its values, and it is precisely the question that the corpus instantiation of Section~6 asks of a parser.

\begin{proposition}[Reliability filters are not monotone]\label{prop:nonmono}
Let $\mathcal{A}_1$ have the states $q_0,q_1,q_2$, the initial state $q_0$, the accepting state $q_2$, and the transitions $q_0\xrightarrow{a}q_1$ and $q_1\xrightarrow{a}q_2$, both weighted $(\tfrac{7}{10},\tfrac{8}{10})$, and let $\mathcal{A}_2$ add a state $p$ with the transitions $q_0\xrightarrow{a}p$, weighted $(1,\tfrac{4}{10})$, and $p\xrightarrow{a}q_2$, weighted $\mathbf{1}$. For $\theta=\tfrac{3}{10}$ and $\rho=\tfrac12$, the word $aa$ belongs to $L_{\theta,\rho}(\mathcal{A}_1)$ and not to $L_{\theta,\rho}(\mathcal{A}_2)$, although $\mathcal{A}_2$ has every run of $\mathcal{A}_1$, whereas the level languages, the supports, and the upward cut-point languages $\{w:\text{value of }w\ge\theta\}$ and $\{w:\text{value of }w>\theta\}$ of scalar max-product automata can only grow when transitions are added.
\end{proposition}

\begin{proof}
In $\mathcal{A}_1$ the only run on $aa$ has weight $(\tfrac{49}{100},\tfrac{16}{25})$, with credible grade $0.3136\ge\theta$ and reliability $0.64\ge\rho$. In $\mathcal{A}_2$ the new run has weight $(1,\tfrac25)$ and credible grade $0.4$, so the Z-sum~\eqref{eq:zoplus} selects it, and its reliability $0.4$ falls below $\rho$. Adding transitions adds runs without removing any, and a maximum over a larger set of runs cannot decrease, which gives the monotonicity of the other languages.
\end{proof}

In the reading of Section~6, the new run is an analysis licensed by a rare construction, of high corpus probability and thin evidence. Once it is available it becomes the most credible analysis of the word, and the reliability filter withdraws the word, reporting that the best analysis now on offer is poorly attested. No scalar max-product weighting produces this behavior under an upward threshold, of the form $\ge\theta$ or $>\theta$, which is the form that admits the words with strong analyses, by the monotonicity just noted. The filter of Proposition~\ref{prop:nonmono} applies to the selected run. An existential filter, which admits a word when \emph{some} accepting run has credible grade at least $\theta$ and reliability at least $\rho$, is monotone under the addition of runs and is regular by the construction in the proof of Proposition~\ref{prop:cut}, and it may suit applications in which any sufficiently reliable analysis is acceptable. The two filters answer different questions, and the selected-run filter is the one that reports on the analysis the device actually returns. The third place is determinization, where, by Proposition~\ref{prop:nondet}, the twins property on the Z-weights is strictly stronger than on their credible grades, so that a Z-automaton may have no deterministic equivalent although its Viterbi image determinizes.

These three phenomena mark the boundary of the conservative extension of Proposition~\ref{prop:transfer}. Below it lies the max-product theory read on the discounted grade; above it lie the reliability filters, their nonmonotonicity, and the determinization condition, which have no counterpart in the credible-grade projection. Both examples were confirmed by exact rational computation.

\subsection{Comparison with other graded devices}\label{ssec:compare}

The results of this section place the Z-devices among the graded devices of the literature, and the comparison is worth making explicit, since several of these devices also carry more than a bare membership degree. Table~\ref{tab:compare} collects the devices with an established decision theory. Max-min fuzzy automata take only finitely many values, because the minimum and the maximum of finitely many grades never produce a new one, so their cut-point languages are regular and their equivalence is decidable~\cite{mordeson2002}. Max-product fuzzy automata, the Viterbi automata of Section~\ref{ssec:weighted}, have regular cut-point languages by Proposition~\ref{prop:cut} applied to weights of the form $\iota(a)=(a,1)$, but their equivalence is undecidable, by the result of Krob~\cite{krob1994} transferred through the encoding of Remark~\ref{rem:cut}. Probabilistic automata sum over runs instead of maximizing, and the sum buys expressive power at the price of tractability: their cut-point languages need not be regular~\cite{rabin1963}, and the emptiness of a cut-point language is undecidable~\cite{paz1971}, although their equivalence is decidable in polynomial time~\cite{tzeng1992}. Intuitionistic fuzzy automata~\cite{jun2005,zhang2009} carry a pair, a degree of membership together with a degree of nonmembership, combined coordinatewise by max-min and min-max rules, so they too take finitely many values; their second component, however, is a further fact about the word, evidence against its membership, and it stands on the same level as the first.

\begin{table}[ht]
\centering\small
\caption{Graded devices with an established decision theory. ``Positive thresholds'' refers to the languages $\{w:\text{value of }w\ge\theta\}$ with $\theta>0$. The decision results assume rational weights and thresholds; for Z-automata the thresholds are on the credible grade and, for the filters, on the reliability of the selected run.}\label{tab:compare}
\begin{tabular}{@{}>{\raggedright\arraybackslash}p{2.6cm}>{\raggedright\arraybackslash}p{2.8cm}>{\raggedright\arraybackslash}p{2.9cm}>{\raggedright\arraybackslash}p{2.9cm}>{\raggedright\arraybackslash}p{3.0cm}@{}}
\toprule
Device & Weight & Competing runs & Positive thresholds & Equivalence \\
\midrule
Max-min fuzzy & grade & maximum of minima & regular & decidable \\
\addlinespace
Max-product fuzzy & grade & maximum of products & regular & undecidable \\
\addlinespace
Probabilistic & probability & sum of products & not regular in general; emptiness undecidable & decidable \\
\addlinespace
Intuitionistic fuzzy & membership and nonmembership & coordinatewise & regular & decidable \\
\addlinespace
Z-automaton & grade and its reliability & most credible run, reported as a pair & regular (Proposition~\ref{prop:cut}) & undecidable; decidable in polynomial time when unambiguous, deterministic included (Proposition~\ref{prop:dec}) \\
\bottomrule
\end{tabular}
\end{table}

Three further ways of carrying a pair through a weighted device are close enough to the Z-weight structure to be confused with it. The componentwise product of two Viterbi semirings, among the separable structures of Proposition~\ref{prop:sep}, maximizes the grade and the reliability independently, so the pair it reports for a word may join the grade of one derivation to the reliability of another and then describes no derivation at all. The lexicographic semirings used to encode sequence models exactly~\cite{roark2011} select by the first component and consult the second only to break ties. The Z-weight structure is itself of this kind: through the injective map $(a,b)\mapsto(ab,a)$ of Section~\ref{ssec:zws} it embeds into the lexicographic product of two Viterbi semirings, ordered by the credible grade and then by the grade. The difference therefore lies in which quantity comes first, not in whether a lexicographic order is used. A lexicographic semiring on the raw pair, ordered by the grade first, lets the reliability only break ties and so can never let it overturn a higher grade. Ordering first by the credible grade lets a well-attested analysis of lower grade defeat a thinly attested one of higher grade, which is the behavior of Example~\ref{ex:parse}. A Pareto structure, which keeps every derivation whose pair is not dominated by another, loses neither component, but it returns a set rather than a weight, so it is not selective and its values grow with the ambiguity of the word. The Z-weight structure $\Zset$ is the selective choice in which the reported pair belongs to a single derivation and the reliability enters the comparison of derivations as a discount on the grade.

The comparison locates the contribution precisely. The Z-devices do not enlarge the class of languages recognized at a threshold, which coincides with that of the max-product devices, and they keep the decidability profile of the max-product devices rather than the richer but less tractable profile of the probabilistic ones. Ordinary scalar Viterbi automata over the discounted weights $a\,b$ already reproduce the primary ranking of the Z-devices (Proposition~\ref{prop:transfer}). What the Z-devices add is a second component that qualifies the first instead of describing the word, kept attached to the derivation it qualifies, so that their additional value lies in the pair-sensitive output and the filters of Section~\ref{ssec:adds}, not in the ranking.

\subsection{What the semiring buys}

Taken together, the results of this section show that the choice of the credible-grade structure in Section~3, vindicated algebraically in Section~4, pays for itself in theory. The rational and the recognizable Z-languages coincide, and they are closed under the rational and the pointwise operations. Their supports are regular, and their level and reliability-filtered languages are regular and effectively constructible. Determinization of unambiguous Z-automata is available under the twins property on the Z-weights, at the price of intermediate weights outside $[0,1]^{2}$, and it meets an obstruction that the Viterbi image does not see (Proposition~\ref{prop:nondet}). The Z-languages sit over the max-product fuzzy languages as a conservative extension that adds the reliability dimension without cost to the existing theory. Among the decision questions considered here, only behavioral equivalence fails to be decidable. It fails not because of the reliability component but because maximization is combined with a product under which finitely many weights generate infinitely many values, as in max-product and tropical automata; max-min fuzzy automata, whose values stay in a finite set, keep a decidable equivalence (Table~\ref{tab:compare}). Equivalence is recovered for unambiguous recognizers, deterministic ones included, a regime that covers some of the evidence-derived recognizers of the next section. The next section turns from this general theory to those constructions, instantiating the Z-automaton with reliabilities estimated from data and examining the Z-languages that result.

\FloatBarrier
\section{Z-languages from data}

The reliability component of a Z-weight was treated abstractly in the preceding sections, as a number in $[0,1]$ attached to a grade, and one way to give it content is to derive it from counts. This section does so for a grammar estimated from a corpus. The grade of a rule is its maximum-likelihood probability, and its reliability is an evidence score that grows with the count of the rule, an attestation penalty rather than a calibrated probability, so that the credible grade of Section~3 becomes the probability of a rule scaled down according to how little it is attested. The section does three things. It shows on a complete example how the attestation penalty can change the selected analysis. It relates the resulting recognizers to the boundary drawn in Section~5: their threshold languages are regular, as they are for every Z-automaton, so only the equivalence of two recognizers is at issue, and it is decidable when the recognizers are unambiguous. Finally, it uses the same evidence-count construction that our applied work used for the reliability of a rating~\cite{turaev2026}.

\subsection{Estimating a Z-grammar from a corpus}

\begin{figure}[t]
\centering
\includegraphics[width=0.8\textwidth]{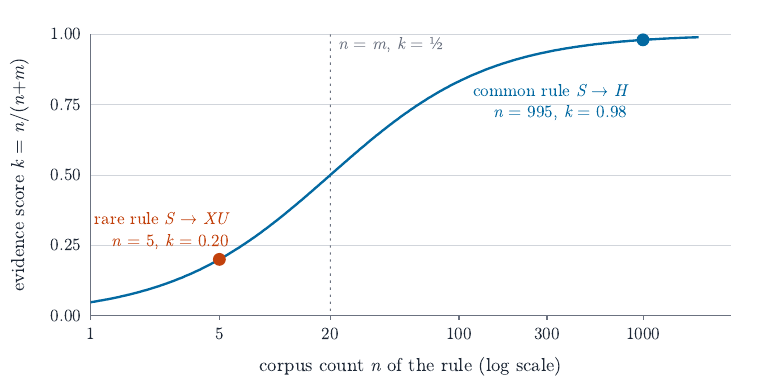}
\caption{The evidence score $k=n/(n+m)$ of an estimated rule against its corpus count $n$, with $m=20$. A rule seen as often as the characteristic count receives one half. The two marked points are the rules $S\to XU$ and $S\to H$ of Example~\ref{ex:parse}. The curve is the definition of the score and is not evidence of its calibration.}\label{fig:corpus}
\end{figure}

Let a grammar be estimated from a parsed corpus in the usual way, by counting how often each production is used. For a nonterminal $X$ with productions $X\to\alpha_{1},\dots,X\to\alpha_{r}$ observed $n_{1},\dots,n_{r}$ times, write $N_{X}=\sum_{j}n_{j}$, and assume $N_{X}>0$. A nonterminal that the corpus never expands receives no estimated productions and so generates nothing, and a production with $n_{j}=0$ is omitted, which amounts to the weight $\mathbf{0}$. We distinguish a \emph{sole allowed} rule, the only production of $X$ that the grammar admits, whose probability is $1$ by definition, from a \emph{sole observed} rule, the only production of $X$ seen in the corpus among several that the grammar admits, whose probability $1$ is a maximum-likelihood estimate from $N_{X}$ observations. The \emph{grade} of $X\to\alpha_{j}$ is its maximum-likelihood probability $\hat p_{j}=n_{j}/N_{X}$, exactly as in a probabilistic grammar, and its \emph{reliability} is the evidence score
\begin{equation}\label{eq:k}
k_{j}=\frac{n_{j}}{n_{j}+m},
\end{equation}
where $m>0$ is a fixed rational \emph{characteristic count}. The score is a modeling choice, an attestation penalty, and we do not claim that it is a calibrated probability. It is motivated by the precision of rare-rule estimates. The relative standard error of a binomial proportion satisfies $\mathrm{RSE}^{2}(\hat p_{j})=(1-p_{j})/(N_{X}p_{j})$, which depends on both $N_{X}$ and $p_{j}$ and is close to $1/n_{j}$ only for rules with a small probability, and~\eqref{eq:k} can be written $k_{j}=1/(1+m\cdot n_{j}^{-1})$, a decreasing function of $1/n_{j}$. The approximation does not cover every rule: for a sole observed rule the binomial error at $\hat p_{j}=1$ vanishes while the count may still be small, and the score then records the number of attestations, not a sampling error. The characteristic count sets the scale, being the count at which the score reaches $1/2$, and it is fixed once for the grammar or once per nonterminal. A Bayesian comparison is limited to a shrinkage coefficient. Under a $\mathrm{Beta}(\alpha,\beta)$ prior with $m=\alpha+\beta$ and mean $\mu=\alpha/m$, the posterior mean of a rule probability from $n_{j}$ occurrences in $N_{X}$ expansions is $[N_{X}/(N_{X}+m)]\,\hat p_{j}+[m/(N_{X}+m)]\,\mu$, whose shrinkage coefficient equals $k_{j}$ when $n_{j}=N_{X}$~\cite{casella1985,efron1973}. This does not make $\hat p_{j}k_{j}$ a posterior mean, which would require both $n_{j}=N_{X}$ and $\mu=0$, and no proper Beta prior with $\alpha>0$ has mean zero. The count $n_{j}$ of the rule, rather than the total $N_{X}$ of its nonterminal, is used so that two rules of one nonterminal with different attestation receive different scores. The Z-weight of the rule is $\zeta(X\to\alpha_{j})=(\hat p_{j},k_{j})$, and its credible grade is $\cred(\hat p_{j},k_{j})=\hat p_{j}k_{j}$, the probability scaled down according to how little the rule is attested. A rule seen once keeps a fraction $1/(1+m)$ of its grade, whereas a rule seen many times keeps nearly all of it. Figure~\ref{fig:corpus} shows the score~\eqref{eq:k} against the count, the hyperbolic curve, sigmoidal on the logarithmic axis, on which a rule seen as often as the characteristic count receives one half, and on which two rules of the example below are marked.

Two properties of the score~\eqref{eq:k} matter for what follows. Its floor is near zero, $k(1)=1/(1+m)$, so a rule seen once is almost entirely discounted, and its ceiling is one, so a rule seen thousands of times is almost not discounted. The rules for which carrying the score matters most are the rarely attested ones, which a probabilistic grammar weights by a probability estimated from a handful of examples. The same kind of score, one that grows with the number of observations behind a value, is the one our applied work used for the ratings of places, where representations with an explicit evidence channel screened out sparsely reviewed candidates~\cite{turaev2026}.

\subsection{Reliability-aware parsing}

The effect of the reliability component is seen most clearly when a sentence is ambiguous and its two analyses rest on evidence of different strength. The credible-grade sum of Section~3 then selects the analysis whose probability best survives the discount for evidence, which is not in general the analysis of highest probability.

\begin{example}[An attestation penalty that changes the selected analysis]\label{ex:parse}
Consider a hypothetical treebank of $10{,}000$ trees and the following estimated productions, with $m=20$; the counts of productions not shown are aggregated into the totals $N_{X}$, and the terminals are $c$ and $d$.
\begin{center}\small
\begin{tabular}{@{}lrrll@{}}
\toprule
Production & $n_{j}$ & $N_{X}$ & grade $\hat p_{j}$ & score $k_{j}$\\
\midrule
$S\to XU$ & $5$ & $10{,}000$ & $0.0005$ & $5/25=0.200$\\
$S\to H$ & $995$ & $10{,}000$ & $0.0995$ & $995/1015\approx0.980$\\
$X\to c$ & $1{,}800$ & $2{,}000$ & $0.9$ & $1800/1820\approx0.989$\\
$H\to WU$ & $995$ & $995$ & $1$ & $995/1015\approx0.980$\\
$W\to c$ & $100$ & $50{,}000$ & $0.002$ & $100/120\approx0.833$\\
$U\to d$ & $400$ & $8{,}000$ & $0.05$ & $400/420\approx0.952$\\
\bottomrule
\end{tabular}
\end{center}
We assume that no production outside the table occurs in a derivation of $cd$, so that the fragment shown is complete for this word. The word $cd$ then has exactly two leftmost derivations, $D_{1}\colon S\Rightarrow XU\Rightarrow cU\Rightarrow cd$ and $D_{2}\colon S\Rightarrow H\Rightarrow WU\Rightarrow cU\Rightarrow cd$. Multiplying the weights of their productions gives
\[
\operatorname{wt}(D_{1})\approx(2.25\times10^{-5},\ 0.188),\qquad \operatorname{wt}(D_{2})\approx(9.95\times10^{-6},\ 0.763),
\]
with credible grades approximately $4.24\times10^{-6}$ and $7.59\times10^{-6}$. The maximum-likelihood parser, reading the grade alone, returns $D_{1}$, whose advantage rests on the production $S\to XU$ seen five times. The credible-grade parser returns $D_{2}$, so that $\Zlang_{G}(cd)=\operatorname{wt}(D_{2})$, although $D_{2}$ is the longer derivation and so bears the length penalty of Remark~\ref{rem:indep}; the production $U\to d$ is common to both derivations and does not affect the comparison. $H\to WU$ is a sole observed rule, whose grade $1$ is an estimate, and its score reflects only its $995$ attestations. The example demonstrates that the proposed attestation penalty can change the selected analysis. Whether this improves parsing accuracy or calibration is an empirical question that is not evaluated here, and the products of scores along $D_{1}$ and $D_{2}$ are declared path scores in the sense of Remark~\ref{rem:indep}, since the productions were estimated from one corpus.
\end{example}

This behavior is a reliability penalty on rare rules, not smoothing in the usual sense, and the difference is worth stating precisely. Smoothing exists chiefly to give unseen rules a nonzero probability, redistributing mass from the seen to the unseen; the Z-grammar does not do this, since an unseen rule has count zero and so credible grade zero, the semiring $\mathbf{0}$. What it does instead is penalize rules that are seen but thinly attested, leaving every estimated probability untouched and carrying alongside it the reliability of its estimate, so that the discount falls not on the probability but on the decision that uses it, and only where the evidence is actually thin. The credible grade $\hat p_{j}k_{j}$ scales the probability of a rule toward nonfiring by a factor that falls as the count of the rule falls.

One point about the product $\hat p_{j}k_{j}$ should be made precise, since collapsing a Z-number into a single reliability-scaled value is in general a lossy conversion. The credible grade orders derivations but is not the value reported: the Z-language returns the full pair of the selected derivation, the product of the grades $\hat p_{j}$ and the product of the scores $k_{j}$ of its rules, and the product serves only to choose among competitors, as a selection semiring requires, its $\otimes$ being preserved only by a multiplicative key. Used this way the reliability scales a grade toward the semiring zero. This resembles the shrinkage rule $\mu+k(\bar{x}-\mu)$ with the target $\mu=0$, but only formally, since a target of zero corresponds to no proper Beta prior and we claim no Bayesian derivation. A cost of the construction is that the discount is toward nonfiring rather than toward a backoff distribution, so a rare rule is discounted toward zero rather than toward a smoothed prior; a reference other than zero is not representable in the semiring, because it is not preserved by the product.

\subsection{Effectivity in the data-driven regime}

Thresholds are effective for every estimated recognizer, since Proposition~\ref{prop:cut} needs no determinism. Some of the finite-state recognizers that arise from estimation are also deterministic or unambiguous, for instance the acceptors underlying morphological analyzers compiled to deterministic transducers, and for these the undecidability result of Section~5 does not apply; the transducers themselves, which produce output, lie outside the present theory (Section~7). Others, such as part-of-speech taggers read as weighted acceptors, are in general neither.

\begin{corollary}[Effectivity for estimated recognizers]\label{cor:eff}
Let the characteristic count $m$ be rational. For every estimated Z-automaton and all rational thresholds $\theta,\rho\in(0,1]$, the level languages $L_{\theta}$ and the reliability-filtered languages $L_{\theta,\rho}$ are regular and effectively constructible. For estimated Z-automata that are deterministic or unambiguous, and for unambiguous estimated right-linear Z-grammars, behavioral equivalence is decidable in polynomial time.
\end{corollary}

\begin{proof}
The estimated weights $(n_{j}/N_{X},\,n_{j}/(n_{j}+m))$ are rational for a rational characteristic count $m$, so the rational-weights hypothesis of Section~5 is satisfied. Regularity is Proposition~\ref{prop:cut}, which requires neither determinism nor unambiguity. A deterministic automaton is unambiguous, and the correspondence of Proposition~\ref{prop:equiv} matches the derivations of a word one to one with its accepting paths, so an unambiguous right-linear Z-grammar yields an unambiguous Z-automaton with the same behavior. Equivalence is then decided in polynomial time by the unambiguous case of Proposition~\ref{prop:dec}.
\end{proof}

The content of the corollary is that the undecidability of behavioral equivalence, which Section~5 inherited from max-product automata and which holds for arbitrary Z-automata, is no obstacle for estimated recognizers that are unambiguous. The acceptor of a morphological analyzer or a chunker compiled to a deterministic finite-state device falls in this regime, and two such acceptors can be compared exactly. Not every estimated sequence model is of this kind. A part-of-speech tagger read as a weighted acceptor is in general neither deterministic nor unambiguous. Its credible-grade thresholds and reliability filters remain effective, but no general procedure decides whether two such taggers define the same Z-language. The corollary thus separates the unambiguous data-driven instances, where every question considered here is decidable, from the ambiguous ones, where only equivalence escapes. The context-free case, in which a parser rather than an acceptor is needed, is among the open problems of Section~7.

\FloatBarrier
\section{Conclusion}\label{sec:concl}

This paper asked whether a reliability attached to a grade, in the manner of the second component of a Z-number, can be carried through a grammar or an automaton by an algebra that keeps the reported pair attached to a single derivation. For scalar grade-reliability pairs the answer is positive, and it has a definite shape. Aggregating the two components separately gives a semiring whose reported pair may belong to no derivation; selecting one derivation keeps the pair coherent but breaks distributivity at degenerate weights; and on the canonical carrier $\{\mathbf{0}\}\cup(0,1]^{2}$, with selection by the credible grade and ties broken by grade, the laws hold. The resulting structure is a commutative selective semiring with a lexicographic description, so the theory of weighted automata over commutative semirings applies to it, and the general results obtained in this way, among them the rational characterization and the closure properties, are corollaries rather than new theorems.

The results that depend on the pair itself are the following. Positive credible-grade cuts and selected-run reliability filters of finite Z-automata with rational weights and thresholds are regular and effectively constructible. A reliability filter on a one-state automaton cannot be implemented by any positive scalar reweighting of the same skeleton, the selected-run filter is not monotone under the addition of runs, and an unambiguous Z-automaton can have no deterministic equivalent although its scalar projection is deterministic, while the twins property suffices for the determinization of unambiguous Z-automata. Among the alternative combination rules, select-by-grade is a semiring on the grade-zero quotient and induces no operation on the double quotient, and the coupled schemes built on the minimum grade law fail distributivity on every carrier examined. Behavioral equivalence is undecidable in general and decidable in polynomial time for unambiguous Z-automata with rational weights.

These results have definite limits. The structure is a model for scalar pairs inspired by Z-numbers and not a semiring of fuzzy-valued Z-numbers, and centroid defuzzification does not recover it from general fuzzy weights (Remark~\ref{rem:fuzzyvalued}). The multiplicative reliability law is a probability only for independent occurrence-level events; for repeated uses of one estimated rule, and for rules estimated from a shared corpus, it is a declared path score, and it penalizes length. The count-based score of Section~6 is an attestation penalty and not a calibrated probability, and the example in which it changes the selected analysis is not an evaluation of parsing accuracy. Whether the twins property suffices for the determinization of ambiguous Z-automata is not proved here.

Relative to the computing-with-words automata, which generalize the input of a fuzzy device, and to type-2 fuzzy grammars, which fuzzify the grade, the Z-devices generalize the weight by a second component that qualifies the first; relative to product and lexicographic semirings, they differ in which quantity orders the derivations (Section~\ref{ssec:compare}). They enlarge no language class and keep the decidability profile of the max-product devices.

Several questions remain open, and we set them out in the order of their likely difficulty.
\begin{enumerate}[leftmargin=*,itemsep=0.3ex,topsep=0.3ex]
\item \textbf{Fuzzy-valued Z-weights.} The theory was developed for scalar pairs. A fuzzy-valued extension is unproved, and centroid recovery fails (Remark~\ref{rem:fuzzyvalued}). Such an extension would need explicit operations on fuzzy grades and reliabilities and an order on fuzzy credible grades compatible with their product, and we do not assume that the comparison of Section~4 would carry over to it unchanged. A decision rule that uses the full fuzzy reliability rather than a summary belongs to the same problem.
\item \textbf{Shared-parameter reliability.} The product law treats the occurrences of rules as independent. A model of reliability for parameters shared across occurrences, and across the rules of one nonterminal estimated from one corpus, would replace the declared path score of Remark~\ref{rem:indep} by a quantity with a probabilistic meaning; it is unlikely to be a semiring product, and it needs a separate model.
\item \textbf{Determinization of ambiguous Z-automata.} Proposition~\ref{prop:det} proves that the twins property suffices for unambiguous Z-automata. Whether it suffices in general, as it does over the tropical semiring, is open.
\item \textbf{The context-free level.} We treated the regular and the recognizable Z-languages. Because $\Zset$ is a commutative semiring, semiring parsing~\cite{goodman1999} already supplies a polynomial-time parsing algorithm for context-free Z-grammars in Chomsky normal form: the CKY recursion evaluated over $\Zset$ returns the most credible parse together with its grade and its reliability. A context-free parser needs, in addition, its own model of reliability for subtrees shared between parses. What remains open is the context-free theory proper. It includes Z-pushdown automata and their equivalence with context-free Z-grammars, normal forms and pumping arguments, and the consequences for Z-languages of the Chomsky--Schützenberger theorem for quantitative context-free languages~\cite{dv2014}, which Section~4 invoked only in passing. Above all, it includes the status of the threshold languages, beginning with whether the level and the reliability-filtered languages of context-free Z-grammars are context-free.
\item \textbf{The weaker weight structures.} The comparison of Section~4 left four strong-bimonoid structures unexploited, among them the counting structures that accumulate rather than select derivations, together with the unclassified native Z-arithmetic. What decision problems these weaker structures can still solve, whether their recognizable and rational Z-languages coincide, and whether a Z-analogue of the probabilistic, sum-over-parses reading of a grammar can be given a usable reliability are questions about the structures that the semiring requirement set aside.
\item \textbf{Learning and smoothing.} The attestation penalty of Section~6 was illustrated, not evaluated. Testing it requires a comparison on a treebank against the established smoothing methods, an estimation of the characteristic count per nonterminal, and a measurement of whether the penalty changes parsing accuracy or calibration.
\item \textbf{Transducers, logic, and trees.} Beyond acceptance and generation lie the Z-weighted transducers that tagging and translation require, the Z-weighted monadic second-order logic whose equivalence with Z-automata the weighted logic-automata connection would predict, and the Z-weighted tree automata that a parse forest, rather than a single parse, would need. Each is a direction in which the construction of this paper is the natural starting point.
\end{enumerate}

\paragraph{Main takeaway and limitations} The main takeaway is that a reliability can be carried through a weighted grammar or automaton by a standard algebra, the credible-grade semiring, provided that the reported pair stays attached to the derivation it qualifies. What the pair then adds over its scalar projection lies in the reported reliability, in the reliability filters, and in determinization, not in the ranking of derivations, which the scalar credible grade already determines. The framework has three practical limitations. It treats scalar pairs rather than fuzzy-valued Z-numbers. Its multiplicative reliability law is a probability only under occurrence-level independence, and otherwise a declared path score that penalizes length. Its usefulness for parsing depends on how reliabilities are obtained from data, which remains to be tested empirically.

\FloatBarrier
\section*{CRediT authorship contribution statement}
\textbf{Sherzod Turaev:} Conceptualization, Methodology, Formal analysis, Investigation, Software, Visualization, Writing -- original draft, Writing -- review \& editing, Project administration. \textbf{Nor Haniza Sarmin:} Validation, Formal analysis, Writing -- review \& editing. \textbf{Wan Heng Fong:} Validation, Writing -- review \& editing.

\section*{Declaration of competing interest}
The authors declare that they have no known competing financial interests or personal relationships that could have appeared to influence the work reported in this paper.

\section*{Funding}
This work was supported by the United Arab Emirates University.

\section*{Data availability}
No data were used for the research described in the article. The scripts that verify the counterexamples and the worked examples by exact rational arithmetic are available from the corresponding author on request.

\section*{Declaration of generative AI and AI-assisted technologies in the manuscript preparation process}
During the preparation of this work the authors used Claude (Anthropic) to assist with language editing and the refinement of draft passages, with the retrieval and verification of bibliographic records, and with the code that draws the figures. The mathematical results, their proofs, and the interpretation of the work are the authors' own. After using this tool, the authors reviewed and edited the content as needed and take full responsibility for the content of the published article.

\section*{Acknowledgements}
The authors gratefully acknowledge the College of Computing and Artificial Intelligence and the Research and Innovation Office at the United Arab Emirates University for the institutional resources, including library access and computational facilities, that supported this work.

\end{document}